\documentclass[%
 aip, 
 amsmath,amssymb,amsthm,
 nofootinbib,
 reprint,
]{revtex4-1}

\usepackage{dcolumn}
\usepackage{bm}
\usepackage[final]{changes}
\usepackage{graphicx}
\usepackage[font=small,labelfont=bf]{caption}
\usepackage{dirtytalk}
\usepackage{csquotes}

\newtheorem{theorem}{Theorem}
\newtheorem{corollary}{Corollary}[theorem]

\begin{document}
\preprint{APS/123-QED}

\title{Variational Principle for Stochastic Mechanics Based on Information Measures}

\author{Jianhao M. Yang}
\email{jianhao.yang@alumni.utoronto.ca}
\affiliation{
Qualcomm, San Diego, CA 92121, USA
}

\date{\today}

\begin{abstract}
Stochastic mechanics is regarded as a physical theory to explain quantum mechanics with classical terms such that some of the quantum mechanics paradoxes can be avoided. Here we propose a new variational principle to uncover more insights on stochastic mechanics. According to this principle, information measures, such as relative entropy and Fisher information, are imposed as constraints on top of the least action principle. This principle not only recovers Nelson's theory and consequently, the Schr\"{o}dinger equation, but also clears an unresolved issue in stochastic mechanics on why multiple Lagrangians can be used in the variational method and yield the same theory. The concept of forward and backward paths provides an intuitive physical picture for stochastic mechanics. Each path configuration is considered as a degree of freedom and has its own law of dynamics. Thus, the variation principle proposed here can be a new tool to derive more advanced stochastic theory by including additional degrees of freedom in the theory. The structure of Lagrangian developed here shows that some terms in the Lagrangian are originated from information constraints. This suggests a Lagrangian may need to include both physical and informational terms in order to have a complete description of the dynamics of a physical system.
\end{abstract}
\maketitle

\section{Introduction}
\label{intro}
Although quantum mechanics is one of the most successful physical theories and has been experimentally confirmed extensively, there are many fundamental questions still left unanswered. For instance, the origin of probability in quantum mechanics is not clearly understood. It is still a curiosity why the probability is calculated as the absolute square of a complex number. The meaning of wave function, especially the interpretation of wave function collapse in a measurement, has been always a debated topic. These questions were not fully addressed by the traditional Copenhagen Interpretation. Over the years in the modern history of quantum physics, many more theories and interpretations have been developed~\cite{Stanford, Jammer74}. 
Among these theories and interpretations, stochastic mechanics is of particular interest because it aims to derive quantum mechanics from classical physics concepts with an additional assumption that a physical system is constantly undergoing a stochastic process~\cite{Goldstein, Nelson2}. There is no need to introduce concepts such as probability amplitude, wave function, or Born's rule, as fundamental elements for the quantum theory. Instead, they are secondary and can be derived. 

\added{Historically, the investigation of the connection between quantum mechanics and diffusion process was started in the early days of quantum mechanics. Schr\"{o}dinger initiated this line of investigation in 1932 by asking a question regarding Brownian motion that is later termed as Schr\"{o}dinger Bridge Problem (SBP)~\cite{SBP}. SBP essentially aims to searching the most likely random evolution that Brownian particles have taken from an initial end point to another end point, provided the probability densities of the two end points are given. In the context of Markov diffusion process, the solution of SBP was found to admit a form that is similar to the Born's rule in quantum mechanics~\cite{Cruz, Leonard}. A more ambitious attempt to formulate quantum mechanics directly based on diffusion process was put forward by Nelson in 1966~\cite{Nelson},} based on an early work due to F\'{e}nyes who recognized the Schr\"{o}dinger equation can be understood as a partial differential equation (PDE) for a Markov process~\cite{Fenyes}. Nelson's theory became the most well-known formulation of stochastic mechanics. In this theory, a rather arbitrary definition of the mean stochastic acceleration was postulated. Then, together with the Fokker-Planck equation, Nelson derived two non-linear PDEs, which when combined together through a set of mathematical transformation, lead to the Schr\"{o}dinger equation. Subsequent researches~\cite{Yasue, Guerra, Pavon} focus on the variational principle that would serves as underlined foundation to recover the Nelson theory. Among them, Yasue's approach~\cite{Yasue} is of importance. It not only successfully derives the stochastic acceleration from the least action principle, but also proposes a generic stochastic calculus for variation. Guerra and Morato~\cite{Guerra} give another variational approach that can recover Nelson' theory, but with a different structure of Lagrangian, leaving a mystery on why multiple Lagrangian can result in the same Nelson's theory.

More recent researches on stochastic mechanics are motivated by two fronts. First, since the source of the randomness that causes the system to perform Brownian motion is not yet completely known, it is desirable to investigate such source. One promising conjecture is that the space metrics itself is stochastic~\cite{Santos,Kurihara}. A particle in such a stochastic space is shown to constantly perform Brownian motion. Given the effect of the randomness of spacetime metrics is universal, the resulting quantum effect is also univeral to all systems in the spacetime. Another proposal, which is less universal since it only applies to charged systems, is that the vacuum electromagnetic field at absolute zero temperature degree is a real radiation. A charged particle constantly interacts with such electromagnetic field. The quantum effects are produced by the electromagnetic noise combined with classical dynamics~\cite{Pena2, Santos2}. Second, in wake of the advance on quantum information, there are considerable amount of interest to explore the informational foundation of quantum mechanics. Information concepts such as entropy, Bayes' Theorem, Fisher information, entanglement, etc., can be considered as foundational elements in constructing quantum theory~\cite{Fuchs, Fuchs02, Frieden, Reginatto, Parwani, Hoehn2014, Yang2017}. Some efforts~\cite{Caticha, Pavon} have been put forward to recover Nelson's theory with entropy as a key element. However, these formulations depend on other assumptions such as conservation of energy~\cite{Caticha} or rather arbitrary constraints in the variation process~\cite{Pavon}.

There are opened questions on how stochastic mechanics can explain quantum phenomena, such as entanglement, locality, the $2\pi$ periodicity of the wave function. Readers are referred to the well known review papers~\cite{Goldstein, Nelson2} and a more recent paper~\cite{Kuipers}. One of the subtleties is that to what extend stochastic mechanics can rely on the classical probability theory to explain quantum mechanics specific phenomena, as exemplified by the multi-time correlations issue~\cite{Grabert, Blanchard}. One should be very cautious in equating quantum mechanics to classical probability theory. Nevertheless, these challenges will continue to inspire future researches to bring new physical insights into stochastic mechanics.

In summary, stochastic mechanics remains a promising theory to explain quantum mechanics classically. The theory can be further developed in multiple fronts. This paper is motivated by exploring possible informational foundation of stochastic mechanics. By doing this we wish to uncover additional physical insights on quantum mechanics. The goal of this work is to take a new look on the variational principle, by examining the structure of Lagrangian and constraints related to information measure. We will show that the efforts are indeed fruitful. By defining proper Lagrangian and imposing constraint of relative entropy for both forward and backward path configurations, we are able to derive the time dynamics for both path configurations using the Yasue stochastic calculus. From them, the Nelson theory and the Schr\"{o}dinger equation are recovered. Furthermore, by adding the Fisher information production into the variational approach, we can also derive the Nelson theory using the Guerra and Morato version of Lagrangian but with the Yasue stochastic calculus. This clears the mystery mentioned earlier. There are several new physical insights our derivation brings in. First, the concept of forward and back paths provides an intuitive physical picture to consider the dynamics of a diffusing particle. It gives more insight on the difference between classical and quantum mechanics in terms of degree of freedom needed to describe a system completely. Second, from methodology perspective, the variational approach presented here can be naturally extended to include new degrees of freedom. It paves a possible way to derive Dirac equation using the stochastic variation method once the theory is formulated in a relativistic setting. Third, The constraint of zero relative entropy imposed to the least action principle shows that the forward and backward paths, although both are needed for a complete description of a diffusing particle, are indistinguishable through measurement. This echoes the idea of interfering alternative proposed in the path integral formulation of quantum mechanics. Lastly, the structure of Lagrangian developed in our works shows that there is intrinsic connection between the physical variables and quantities related information measure. It is subtle to distinguish them when writing down a Lagrangian that consists many terms. This suggests that it is possible to reconsider from the information measure perspective the meaning of certain terms in the Lagrangian density used in classical or quantum field theory.

Although the formulation presented here is mathematically equivalent to the Nelson theory, we believe the method and the conceptual insights brought up in this work can be valuable for future investigation on the foundation of quantum mechanics.

The paper is organized as followings. Section \ref{sec:concept} briefly reviews Nelson's theory and the stochastic calculus of variation proposed by Yasue. In Section \ref{sec:results} we present the main results. An information measure, the relative entropy for the froward and backward path configurations, is introduced. We then prove that for a Markov diffusion process, the relative entropy must be zero. Using this as a constraint in the least action principle, we derive the Nelson theory in Section \ref{subsec:var1}. In Section \ref{subsec:var2}, another information measure, the Fisher information, is introduced. This allows us to recover Nelson's theory using Guerra and Morato version of Lagrangian. Section \ref{subsec:Lagrangian} compares three sets of Lagrangian that can be used to derive Nelson's theory when coupled with proper information measures. In Section \ref{sec:conclusion}, we discuss the physical insights brought in by our formulation, point out the limitations of our derivation, and summarize the conclusions. 

\section{Stochastic Mechanics and Variation}
\label{sec:concept}
This section briefly reviews the stochastic mechanics and emphasizes on the formulation that is referred in later sections. For convenience we will adopt the mathematical notations in Nelson's works ~\cite{Nelsonbook}.

\subsection{Nelson's Theory}
\label{subsec:definition}
The basic assumption for the stochastic mechanics is to consider a system as a point particle and constantly undergoes a Brownian motion. Let $\xi(t)$ be a Markov diffusion process\footnote{We limit the study here to Markov diffusion process only, even though part of Nelson's formulation can be app
lied to non-Markov process. Note that the Brownian motion here is an energy conservation process, rather than a classical dissipative diffusion, see the detailed explanation in section 14 of Nelson's book~\cite{Nelsonbook}.} such that
\begin{equation}
\label{fwd}
    d\xi^i(t) = b^i_+(\mathbf{\xi}(t), t)dt + dW^i_+(t)
\end{equation}
where $i=1,2,3$ is the spatial index. $\mathbf{b}_+(\mathbf{\xi}(t),t)$ is a vector-valued function which meaning will be given shortly. $W_+^i(t)$ is the standard, independent Wiener process with properties $E[dW_+^i(t)] =0$ and $E[dW^i_+(t)dW^j_+(t)] = 2\nu\delta^{ij}dt$, where $E[\cdot]$ denotes the absolute expectation value. The diffusion coefficient $\nu$ is set to be inversely proportional to the mass of the system $\nu = \hbar/2m$ and $\hbar$ is determined later.

Given that the stochastic process $\mathbf{\xi}(t)$ is not differentiable, the forward and backward derivatives are defined to replace the regular derivative. For a real-valued stochastic process $f(\mathbf{\xi}(t))$ (also denoted as $F(t)$ for simpler notation), its forward derivative $D_+$ and backward derivative $D_-$ are defined as~\cite{Nelson}
\begin{equation}
\label{Derivative}
\begin{split}
    D_+F(t) &= \lim_{dt\to 0^+} E_t [\frac{F(t+dt)-F(t)}{dt}] \\
    D_-F(t) &= \lim_{dt\to 0^+} E_t [\frac{F(t)-F(t-dt)}{dt}] \\
\end{split}
\end{equation}
where $E_t[\cdot]$ is a conditional expectation operator with respect to the configuration at time $t$. More precisely, the conditional expectation should be denoted as $E_{\xi(t)}[\cdot]$, we use $E_t[\cdot]$ for simpler notation. For instance, given $t'\ne t$, 
\begin{equation}
\label{Et}
    E_t[f(\mathbf{\xi}(t))] = \int f(\mathbf{\xi}'(t'))p(\mathbf{\xi}',t'|\mathbf{\xi},t)d\mathbf{\xi}'
\end{equation}
where $p(\mathbf{\xi}',t'|\mathbf{\xi},t)$ is the conditional probability density. With this definition, it becomes clear that $D_+\xi^i(t) = b^i_+(\mathbf{\xi}(t),t)$ is the mean forward velocity. The diffusion process can also be written as 
\begin{equation}
\label{bwd}
    d\xi^i(t) = b^i_-(\mathbf{\xi}(t), t)dt + dW^i_-(t)
\end{equation}
where $dW^i_-(t)$ has the same property as $dW^i_+(t)$ except it is independent of $\xi(s)$ for $s\geq t$. By the definition in (\ref{Derivative}), $D_-\xi^i(t) = b^i_-(\mathbf{\xi}(t),t)$, which is the mean backward velocity. With the definitions (\ref{fwd}), (\ref{Derivative}), (\ref{bwd}), one obtains the following explicit expressions
\begin{equation}
    \label{Df}
    \begin{split}
    D_+f(\mathbf{\xi}(t), t) &= (\partial /\partial t + {b_+^i}\nabla_i +\nu\Delta)f(\mathbf{\xi}(t), t) \\
    D_-f(\mathbf{\xi}(t), t) &= (\partial /\partial t + {b_-^i}\nabla_i -\nu\Delta)f(\mathbf{\xi}(t), t)
    \end{split}
\end{equation}
Let $\rho(\mathbf{x},t)$ be the probability density of the diffusion process at position $\mathbf{x}$ at time $t$. Nelson's theory assumes $\rho(\mathbf{x},t)$ satisfies both the forward and backward Fokker-Planck equations. From the forward and backward Fokker-Planck equations, the continuity equation for $\rho(\mathbf{x},t)$ is obtained.
\begin{equation}
    \label{contP}
    \frac{\partial \rho}{\partial t} = - \frac{1}{2}\nabla_i((b_+^i + b_-^i)\rho).
\end{equation}
More crucially, the following identity is also derived,
\begin{equation}
    \label{osmotic}
    b_+^i - b_-^i = 2\nu\nabla^i ln(\rho)
\end{equation}
A more elegant derivation of (\ref{osmotic}) is given based on Bayes' theorem~\cite{Caticha} 
\begin{equation}
    \label{Bayes}
    p(\mathbf{x}'|\mathbf{x}) = \frac{p(\mathbf{x}|\mathbf{x}')\rho(\mathbf{x}')}{\rho(\mathbf{x})},
\end{equation}
and the definition of (\ref{Derivative}), and (\ref{Et}) by taking $\mathbf{x}=\mathbf{\xi}(t)$ and $\mathbf{x}'=\mathbf{\xi}'(t')$. This approach has its advantage since Bayes' theorem is more general in probability theory and there is no need to depend on the Fokker-Planck equation.

The drift velocity is defined as $v^i=(b_+^i + b_-^i)/2$, and the so-called ``osmotic velocity", which is somewhat misleading and will be further discussed later, is defined as $u^i=(b_+^i - b_-^i)/2 = \nu\nabla^i ln\rho$. Multiplying $1/\rho$ to both sides of the continuity equation (\ref{contP}), and taking the gradient of both sides, one gets the Nelson first equation
\begin{equation}
    \label{Nelson1}
    \frac{\partial u^i}{\partial t} = -\nu\Delta v^i - \nabla^i (v^ju_j).
\end{equation}
The mean acceleration is defined somewhat arbitrarily as
\begin{equation}
    \label{acc}
    a^i = \frac{1}{2}(D_+b_-^i + D_-b_+^i).
\end{equation}
Assuming Newton's second law holds $\mathbf{F} = m\mathbf{a} = -\nabla\phi$, where $\phi$ is the external potential, and applying (\ref{Df}) to $b_-^i$ and $b_+^i$ in (\ref{acc}), lead to the Nelson's second equation
\begin{equation}
    \label{Nelson2}
    \frac{\partial v^i}{\partial t} = (u^i\nabla_i)u^i-(v^i\nabla_i)v^i+\nu\Delta u^i - \frac{\nabla^i\phi}{m}.
\end{equation}
Eqs. (\ref{Nelson1}) and (\ref{Nelson2}) give the complete description of the dynamics of a Brownian particle in the context of stochastic mechanics. By introducing a series of mathematical transformations, Eqs. (\ref{Nelson1}) and (\ref{Nelson2}) can be combined into a single linear PDE with complex variable. Let $\nabla^iR=\frac{m}{\hbar}u^i$, $\nabla^i S=\frac{m}{\hbar}v^i$, and $\psi=e^{R+iS}$, one can verify that (\ref{Nelson1}) and (\ref{Nelson2}) are equivalent to
\begin{equation}
    \label{SE}
    i\hbar\frac{\partial \psi}{\partial t} = (-\frac{\hbar^2}{2m}\Delta + \phi)\psi,
\end{equation}
which is the Schr\"{o}dinger equation. Given $\nabla^iR=u^i/(2\nu)$ and $u^i = \nu\nabla^i ln\rho$, we have $\rho = e^{2R}$. Thus,
\begin{equation}
    \label{Born}
    \rho = |\psi|^2 = e^{2R}.
\end{equation}
The Born's rule is naturally derived rather than being postulated. 

The description of a diffusing particle in stochastic mechanics differs from the classical mechanics in that it requires two non-linear PDEs, rather than just one. Extra degree of freedom is introduced through the forward and backward velocities. Note that (\ref{Nelson1}) is essentially derived from the Fokker-Planck equations for $\rho$ and the identity (\ref{osmotic}). This indicates that a component of Schr\"{o}dinger equation comes from the probability theory itself, which we will explore later in terms of information quantity. For now, we turn back to the definition of mean acceleration (\ref{acc}), which is rather arbitrary. It is desirable to justify (\ref{acc}) from a first principle. This motivates the development of a variational approach. 

\subsection{Stochastic Calculus of Variations}
\label{calculus}
There are multiple variational methods proposed to recover Nelson's theory based on the least action principle~\cite{Yasue, Guerra, Pavon} or the conserved energy constraint~\cite{Caticha}. The Yasue's variation method is of particular interest here, since we will extensively use its stochastic calculus, which we give a brief overview here. More rigorous description of the stochastic calculus can be found in Ref.~\cite{Yasue}.

In Yasue's stochastic calculus, the three variables, $(\mathbf{\xi, b_+, b_-})$, or equivalently, $(\mathbf{\xi}, D_+\mathbf{\xi}, D_-\mathbf{\xi})$, are considered independent during the variation process. Thus, the Lagrangian is denoted as $L(\mathbf{\xi}, D_+\mathbf{\xi}, D_-\mathbf{\xi})$. Suppose a particle moves from point $a$ (i.e., $\mathbf{\xi}(t_a)=\mathbf{x}_a$) to point $b$ (i.e., $\mathbf{\xi}(t_b)=\mathbf{x}_b$), the stochastic action is defined as 
\begin{equation}
    \label{action}
    J_{ab}=E[\int_{t_a}^{t_b} L(\mathbf{\xi}, D_+\mathbf{\xi}, D_-\mathbf{\xi}) dt].
\end{equation}
Here $E[\cdot]$ denotes the absolute expectation along the path $\mathbf{\xi}(t_a)\to\mathbf{\xi}(t_b)$ and $t_a < t_b$. Now let the path $\xi$ vary but keep the end points fixed as $\mathbf{x}_a$ and $\mathbf{x}_b$. The variation itself is a stochastic process, denoted as $\mathbf{z}(t)$ for $t_a<t<t_b$ and $\mathbf{z}(t_a)=\mathbf{z}(t_b)=0$. The variation of stochastic action, due to the variation of $\xi$, is $\delta J_{ab}=J_{ab}(\mathbf{\xi+z}) - J_{ab}(\mathbf{\xi})$. Let $\|\cdot\|$ denote the Euclidean vector norm. Defining
\begin{equation}
    \|\mathbf{z}\| := \sup_{t\in [t_a, t_b]}( \|\mathbf{z}(t)\| + \|D_+\mathbf{z}(t)\| + \|D_-\mathbf{z}(t)\|),
\end{equation}
we say $\xi(t)$ is critical for $J_{ab}$ if $\delta J_{ab}=o(\|\mathbf{z}\|)$ holds for all time interval $[t_a, t_b]$ and for arbitrary variation $\mathbf{z}(t)$, and with the condition that $J_{ab}$ has finite energy. To obtain the explicit expression of $\delta J_{ab}$, one performs Taylor's expansion of $L(\mathbf{\xi}, D_+\mathbf{\xi}, D_-\mathbf{\xi})$,
\begin{equation}
\label{dJ1}
    \delta J_{ab}=E[\int_{t_a}^{t_b}\{\frac{\partial L}{\partial \xi^i}z^i + \frac{\partial L}{\partial D_+\xi^i}D_+z^i + \frac{\partial L}{\partial D_-\xi^i}D_-z^i\}dt].
\end{equation}
To proceed further, we need the following identity~\cite{Nelsonbook} for stochastic processes $f(t)$ and $z^i(t)$
\begin{equation}
    \label{productrule}
    \begin{split}
        \int_{t_a}^{t_b} E[f(t)D_-z^i(t)]dt =& - \int_{t_a}^{t_b} E[z^i(t)D_+f(t) ]dt \\
        &+ E[f(t)z^i(t)|^{t_b}_{t_a}],
    \end{split}
\end{equation}
Due to the property of Markov process, the expectation operator $E[\cdot]$ and the integration over $t$ can exchange the order (see Appendix B). This allows us to apply (\ref{productrule}) to (\ref{dJ1}) by taking $f(t)=\partial L/\partial D_{\pm}\xi^i$. Recalling $z(t_a)=z(t_b)=0$, and replacing $D_+\xi^i=b^i_+$ and $D_-\xi^i=b^i_-$, we obtain
\begin{equation}
\label{dJ3}
    \delta J_{ab}=E[\int_{t_a}^{t_b}\{\frac{\partial L}{\partial \xi^i} - D_-( \frac{\partial L}{\partial b^i_+}) - D_+(\frac{\partial L}{\partial b_-^i})\}z^idt].
\end{equation}
Since $\mathbf{z}(t)$ is an arbitrary stochastic process, $\delta J_{ab}=0$ if and only if 
\begin{equation}
    \label{EL-eq}
    \frac{\partial L}{\partial \xi^i} - D_-( \frac{\partial L}{\partial b^i_+}) - D_+(\frac{\partial L}{\partial b_-^i}) =0.
\end{equation}
Defining the Lagrangian for the diffusion process $\xi(t)$ as
\begin{equation}
    \label{L_Y}
    L_Y(\mathbf{\xi, b_+, b_-}) = \frac{1}{2}[\frac{1}{2}m(b_+)_ib_+^i + \frac{1}{2}m(b_-)_ib_-^i] - \phi(\xi),
\end{equation}
and substituting it into (\ref{EL-eq}), we have
\begin{equation}
\label{YasueEq}
    \frac{1}{2}m(D_+b_-^i + D_-b_+^i) + \nabla^i\phi = 0.
\end{equation}
This justifies the definition of mean acceleration in (\ref{acc}).

The choice of Lagrangian in (\ref{L_Y}) appears to be intuitive as it assumes the kinetic energy is the average of kinetic energy associate with forward and backward velocities. However, there are multiple variables related to velocities, such as $\mathbf{b_+}, \mathbf{b_+}, \mathbf{v}$ and $\mathbf{u}$. There is flexibility to define the kinetic energy and consequently the Lagrangian. For instance, Guerra and Morato suggested a different Lagrangian~\cite{Guerra} as
\begin{equation}
    \label{L_G}
    L_G(\mathbf{\xi, b_+}) = \frac{1}{2}m(b_+)_ib_+^i + \frac{\hbar}{2}\nabla_ib_+^i - \phi(\xi).
\end{equation}
Using $L_G$ with a different variation approach, they are able to recover Nelson's theory as well. It has been a mystery why two different Lagragians lead to the same result~\footnote{Nelson stated~\cite{Nelsonbook} that ``We have seen this is so, but I wish I understood why it is so."}. We will later show that both versions of Lagrangian can be unified in our variation method. Another interesting point is that the Lagrangian in (\ref{L_Y}) indicates that the Lagrangian can be split into two parts, one for forward path configuration and the other for backward path configuration. One might ask if there is dynamics equation that can be derived for each of the path configuration using the variation calculus. These are interesting questions to be answered next. 

\section{Results}
\label{sec:results}
\subsection{Relative Entropy in Path Space}
\label{subsec:relH}
To search the dynamics equations for the forward and backward path configurations, we first introduce the relative entropy of the forward and backward paths, as this in the end leads to a constraint in the stochastic variational principle we will propose. Note that the terms forward or backward do not refer to the direction of time. Instead, it just refers to a path configuration that is described by either the forward mean velocity or the backward mean velocity.

\begin{figure*}
\begin{center}
\includegraphics[scale=3.9]{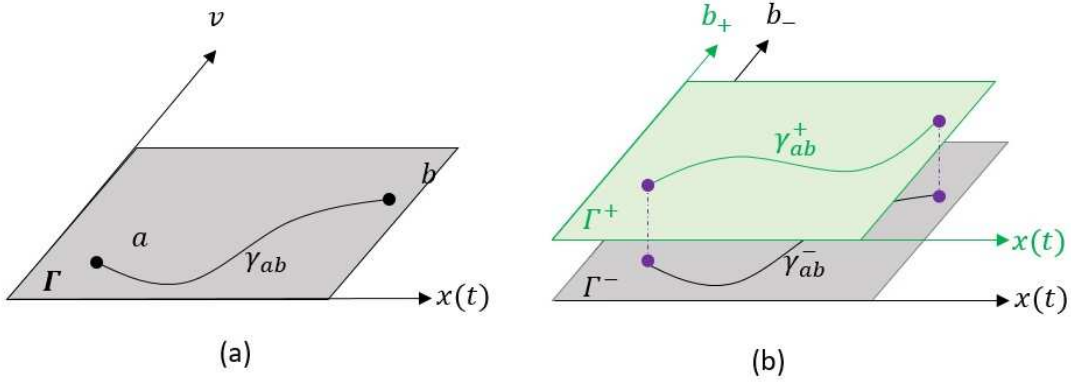}
\caption{Plots of trajectory path of a point particle in phase space. (a) In classical mechanics, a point particle moves from point $a$ to point $b$ with a path configuration $\gamma_{ab}$ determined by least action principle. (b) In stochastic mechanics, diffusion of a point particle is described with forward and backward path configurations $\gamma^{\pm}_{ab}$ in the respective phase space $\Gamma^{\pm}$.}
\label{fig:1}       
\end{center}
\end{figure*}

Suppose a particle undergoes a Markov diffusion process $\mathbf{\xi}(t)$ from $t_a\to t_b$. We divide the time interval $[t_a, t_b]$ into $t_1=t_a, t_2=t_1+\Delta t, ..., t_n=t_b=t_1+n\Delta t$, and the corresponding local position variables are $\mathbf{x}_1, \mathbf{x}_2, ..., \mathbf{x}_n$.  Configuration $\gamma^+_{ab}: t\in [t_a, t_b] \mapsto \gamma^+_{ab}(t) = (\mathbf{x}(t), \mathbf{b_+}(\mathbf{x}(t))$ is defined as a forward sample path from $\mathbf{x}_a$ to $\mathbf{x}_b$, while configuration $\gamma_{ab}^-: t\in [t_a, t_b]\mapsto \gamma^-_{ab}(t) = ( \mathbf{x}(t), \mathbf{b_-}(\mathbf{x}(t))$ is defined as a backward sample path. It is important to note that $\gamma^+_{ab}$ and $\gamma^-_{ab}$ share the same position variable $\mathbf{x}(t)$ but differ by the velocity variables. If one plots the trajectory paths of $\gamma^+_{ab}$ and $\gamma^-_{ab}$ in the four dimension space-time, both paths overlap. But if one plots the paths in the phase spaces defined next, they are different. In the phase space for a diffusing point particle, the particle is identified by $\Gamma: \{(\mathbf{x}, \mathbf{b}_+, \mathbf{b}_-)\}$. We can envision the phase space $\Gamma$ comprises two classical phase subspaces $\Gamma^+: \{(\mathbf{x}, \mathbf{b}_+)\}$ and $\Gamma^-: \{(\mathbf{x}, \mathbf{b}_-)\}$. Plotting $\gamma^+_{ab}$ in $\Gamma^+$ and $\gamma^-_{ab}$ in $\Gamma^-$ gives two different trajectory paths, as shown in Fig.\ref{fig:1}b.

The probability density within a space measure ${\cal{D}}\mathbf{x} = d\mathbf{x}_1d\mathbf{x}_2...d\mathbf{x}_n$ is $\rho(\mathbf{x}_1, t_1; \mathbf{x}_2, t_2; ..., \mathbf{x}_n, t_n)$.  In other words, the regular probability measure on the path space $dP = \rho{\cal{D}}\mathbf{x}$. By the Markov properties, the probability density for the forward path can be written as\footnote{In the equation, we suppress the time variable $t_1, t_2, ..., t_n$ without lost of clarity since the subscript index $i$ also serves as parameter showing the sequence of time. Thus, the transition probability $p(\mathbf{x}_i|\mathbf{x}_{i-1})$ should be understood as the standard notation of $p(\mathbf{x}_i, t_i|\mathbf{x}_{i-1}, t_{i-1})$. This notation convention will be used throughout the paper.}
\begin{equation}
    \label{fwdp}
    \rho_+(\mathbf{x}_1, ..., \mathbf{x}_n)=\rho(\mathbf{x}_1)p(\mathbf{x}_2|\mathbf{x}_1)p(\mathbf{x}_3|\mathbf{x}_2)...p(\mathbf{x}_n|\mathbf{x}_{n-1}),
\end{equation}
so that $dP_+ = \rho_+{\cal{D}}\mathbf{x}$. Similarly the probability density for the backward  path is
\begin{equation}
    \label{bwdp}
    \rho_-(\mathbf{x}_1, ..., \mathbf{x}_n)=p(\mathbf{x}_1|\mathbf{x}_2)p(\mathbf{x}_2|\mathbf{x}_3)...p(\mathbf{x}_{n-1}|\mathbf{x}_{n})\rho(\mathbf{x}_n),
\end{equation}
and $dP_- = \rho_-{\cal{D}}\mathbf{x}$. The relative entropy of forward path probability to the backward path probability, i.e., the  Kullback–Leibler divergence, is
\begin{equation}
    \label{fwdH}
    H(\rho_+\|\rho_-) = \int dP_+ ln(\frac{dP_+}{dP_-}) = \int \rho_+ln(\frac{\rho_+}{\rho_-}){\cal{D}}\mathbf{x}.
\end{equation}
Similarly,
\begin{equation}
    \label{bwdH}
    H(\rho_-\|\rho_+) = \int dP_- ln(\frac{dP_-}{dP_+}) = \int \rho_-ln(\frac{\rho_-}{\rho_+}){\cal{D}}\mathbf{x}.
\end{equation}
The following theorem gives explicit expression of the relative entropy.
\begin{theorem} 
\label{relativeH}
For a Markov diffusion process $\mathbf{\xi}(t)$ from $t_a\to t_b$, the relative entropy can be written as
\begin{equation}
\label{theo1}
\begin{split}
     H(\rho_+\|\rho_-) &= H_b - H_a + \int_{t_a}^{t_b} E[D_+ln(\rho(\mathbf{\xi}(t))]dt \\
     H(\rho_-\|\rho_+) &= H_a - H_b - \int_{t_a}^{t_b} E[D_-ln(\rho(\mathbf{\xi}(t))]dt, 
\end{split}
\end{equation}     
where $\rho(\mathbf{x}(t))$ is the probability density at $\mathbf{x}(t)$, and 
\begin{equation}
\begin{split}
    H_a &= -\int \rho(\mathbf{x}_a)ln(\rho(\mathbf{x}_a))d\mathbf{x}_a,\\ H_b &= -\int \rho(\mathbf{x}_b)ln(\rho(\mathbf{x}_b))d\mathbf{x}_b.
\end{split}
\end{equation}
\end{theorem}
Applying (\ref{osmotic}) to (\ref{theo1}) leads to Corollary 1.1. 
\begin{corollary}
\label{cor1}
An alternative expression of the relative entropy is
\begin{equation}
\label{theo1_2}
\begin{split}
     H(\rho_+\|\rho_-) &= H_b - H_a - \frac{1}{2}\int_{t_a}^{t_b} E[\nabla_i(b_+^i + b_-^i)]dt \\
     H(\rho_-\|\rho_+) &= H_a - H_b + \frac{1}{2}\int_{t_a}^{t_b} E[\nabla_i(b_+^i + b_-^i)]dt.
\end{split}
\end{equation}
\end{corollary}
Proof of the Theorem and the Corollary is given in Appendix \ref{AppendixA}. One can immediately observe that $H(\rho_+\|\rho_-)=-H(\rho_-\|\rho_+)$. Then, by the non-negativity property of relative entropy~\cite{Nelson00}, we have
\begin{corollary}
\label{cor2}
$H(\rho_+\|\rho_-) = H(\rho_-\|\rho_+) = 0$.
\end{corollary}
Since $H(\rho_+\|\rho_-)=0$ if and only if $\rho_+=\rho_-$, the two path probability densities are the same, which is consistent with the results in Ref.~\cite{Nelsonbook}. 

Corollary \ref{cor2} gives two constraints in terms of relative entropy. These two constraints in fact are the same as the transport equations~\cite{Guerra}. The significance of Corollary \ref{cor2} is that we recognize them as relative entropy. It says that the information encoded in the forward path configuration and backward path configuration is identical. The dynamics of the Brownian motion needs to comply to this constraint. If we define the total relative entropy as $H_{rel}=H(\rho_+\|\rho_-) + H(\rho_-\|\rho_+)$, by Theorem \ref{relativeH}, $H_{rel}=0$ and doesn't result in a constraint. We will next see how these constraints play important roles in the stochastic variation.

\subsection{Variations with Relative Entropy Constraint}
\label{subsec:var1}
In this subsection, we will show how the dynamic equations for the forward and backward mean velocities are derived through variational method. From them the Schr\"{o}dinger equation is recovered.

The Yasue version of Lagrangian (\ref{L_Y}) already suggests we can split it into two parts, forward and backward Lagrangian. Let
\begin{equation}
\label{LYdir}
\begin{split}
    L_{Y}^+(\mathbf{\xi, b_+}) &= \frac{1}{2}m(b_+)_ib_+^i - \phi(\mathbf{\xi}), \\
    L_{Y}^-(\mathbf{\xi, b_-}) &= \frac{1}{2}m(b_-)_ib_-^i - \phi(\mathbf{\xi}).
    \end{split}
\end{equation}
Then, $L_{Y}$ defined in (\ref{L_Y}) is $L_Y=(L_{Y}^++L_{Y}^-)/2$. The corresponding forward and backward actions are defined as
\begin{equation}
    \label{YAction}
    \begin{split}
    A_{ab}^+ &= E[\int_{t_a}^{t_b} L_{Y}^+dt] \\
    A_{ab}^- &= E[\int_{t_a}^{t_b} L_{Y}^-dt],
    \end{split}
\end{equation}
where $E[\cdot]$ denotes the absolute expectation along the path $\mathbf{x}_a \to \mathbf{x}_b$. In Appendix \ref{AppendixB}, we show that for a Markov diffusion process, one can swap the order of integration and taking expectation. Thus,
\begin{equation}
    \label{YAction2}
    \begin{split}
    A_{ab}^+ &= \int_{t_a}^{t_b} E[L_{Y}^+]dt \\
    A_{ab}^- &= \int_{t_a}^{t_b} E[L_{Y}^-]dt,
    \end{split}
\end{equation}
and $E[\cdot]$ denotes the absolute expectation at time $t$. Now we combine the forward and backward actions with the constraints in Corollary \ref{cor2} by defining the Lagrangian functional
\begin{equation}
\label{LF_Y}
\begin{split}
    J_{ab}^+ &= A_{ab}^+ - \beta H(\rho_+\|\rho_-), \\
    J_{ab}^- &= A_{ab}^- - \beta H(\rho_-\|\rho_+), 
\end{split}
\end{equation}
where $\beta$ is a Lagrangian multiplier. The definition of Lagrangian functional means that we seek to minimize the forward and backward actions with the relative entropy constraints. Substitute (\ref{theo1_2}) and (\ref{LYdir}) into (\ref{LF_Y}),
\begin{equation}
\label{LF_Y2}
    \begin{split}
    J_{ab}^+ =& \int_{t_a}^{t_b} E[\frac{1}{2}m(b_+)_ib_+^i - \phi(\xi)  +\frac{\beta}{2}\nabla_i(b_+^i + b_-^i)]dt\\
    & -\beta(H_b - H_a).
    \end{split}
\end{equation}
Next we apply the stochastic calculus described in Section \ref{calculus}. Suppose we variate the diffusion process $\mathbf{\xi}(t)$ to be $\mathbf{\xi'}(t)=\mathbf{\xi}(t)+\mathbf{z}(t)$ with $\mathbf{z}(t_a)=\mathbf{z}(t_b)=0$, and demands that $\delta J_{ab}^+=o(\|\mathbf{z}\|)$. Substituting (\ref{LF_Y2}) into (\ref{dJ1}), and notice that $H_a$ and $H_b$ are fixed values, we have
\begin{equation}
\label{J2}
    \begin{split}
    \delta J_{ab}^+ &= \int_{t_a}^{t_b} E[\delta L_{Y}^+]dt + \frac{\beta}{2}\int_{t_a}^{t_b} E[\nabla_i(\delta b_+^i + \delta b_-^i)]dt\\
    & = \int_{t_a}^{t_b} E[m(b_+)_i\delta b_+^i -(\nabla_i\phi) z^i + \frac{\beta}{2}\nabla_i(\delta b_+^i + \delta b_-^i)]dt \\
    \end{split}
\end{equation}
To proceed further, we need the following identity that is derived from integration by part. Let $\mathbf{f}$ a smooth vector function of the diffusion process,
\begin{equation}
\label{id}
\begin{split}
    E[\nabla_i f^i]& =\int \rho\nabla_i f^i d\mathbf{x} 
    = - \int (\nabla_i\rho)f^i d\mathbf{x} \\
    &= - E[(\nabla_i ln(\rho)) f^i] = -\frac{1}{\nu}E[u_if^i].
    \end{split}
\end{equation}
Let $f^i=\delta b_+^i$, we have $E[\nabla_i \delta b_+^i]=- E[\delta b_+^i\nabla_i ln(\rho)]$. Substituting it into (\ref{J2}), and replacing $\delta b_+^i=D_+z^i$, $\delta b_-^i=D_-z^i$ (see proof in Appendix \ref{AppendixC}), we get
\begin{equation}
\label{J4}
    \begin{split}
    \delta J_{ab}^+ =& \int_{t_a}^{t_b} E[m(b_+)_i\delta b_+^i -(\nabla_i\phi) z^i \\
    &- \frac{\beta}{2} (\delta b_+^i + \delta b_-^i)\nabla_i ln(\rho)]dt \\
    =&\int_{t_a}^{t_b} E[m(b_+)_iD_+z^i -(\nabla_i\phi) z^i \\
    &-\frac{\beta}{2} (D_+z^i + D_-z^i)\nabla_i ln(\rho)]dt 
    \end{split}.
\end{equation}
Applying (\ref{productrule}) to (\ref{J4}), we obtain
\begin{equation}
\label{J5}
    \delta J_{ab}^+ = -\int_{t_a}^{t_b} E[\{mD_-b_+^i +\nabla^i\phi - \frac{\beta}{2}(D_+ + D_-)\nabla^i ln(\rho)\}z_i]dt.
\end{equation}
Since $z_i(t)$ is arbitrary diffusion process, $\delta J_{ab}^+=0$ if and only if
\begin{equation}
    \label{fwdPDE}
    mD_-b_+^i +\nabla^i\phi - \frac{\beta}{2}(D_+ + D_-)\nabla^i ln(\rho) =0.
\end{equation}
This is the PDE for the dynamics derived from the forward path. Repeat the same variational method on $J_{ab}^-$, we obtain the PDE for the dynamics of the backward path,
\begin{equation}
    \label{bwdPDE}
    mD_+b_-^i +\nabla^i\phi + \frac{\beta}{2}(D_+ + D_-)\nabla^i ln(\rho) =0.
\end{equation}
(\ref{fwdPDE}) + (\ref{bwdPDE}) results in $mD_-b_+^i + mD_+b_-^i +2\nabla^i\phi = 0$ which is the same as (\ref{YasueEq}) and leads to Nelson's second equation (\ref{Nelson2}). (\ref{fwdPDE}) - (\ref{bwdPDE}) gives
\begin{equation}
    \label{CombinedPDE}
    mD_-b_+^i - mD_+b_-^i = \beta(D_+ + D_-)\nabla^i ln(\rho).
\end{equation}
In Appendix \ref{AppendixD}, we prove that if $\beta=\hbar$, (\ref{CombinedPDE}) is the same as Nelson's first equation (\ref{Nelson1}). Note that Nelson's equations are written in terms of time dynamics of $v^i$ and $u^i$. Here however, using (\ref{osmotic}) and (\ref{Df}), we can express the time dynamics equation (\ref{bwdPDE}) in terms of $b_{\pm}^i$ as following,
\begin{equation}
    \label{fwddynamics}
    \begin{split}
       \frac{\partial b_+^i}{\partial t} =& \frac{1}{2}[-(b_+^j\nabla_j + b_-^j\nabla_j)b_+^i- (b_+^j\nabla_j \\
       &-b_-^j\nabla_j) b_-^i] - \frac{\hbar}{2m}\Delta b_-^i -\frac{\nabla^i\phi}{m}.
    \end{split}
\end{equation}
Similarly the time dynamics of backward mean velocity is derived from (\ref{fwdPDE}) as
\begin{equation}
    \label{bwddynamics}
    \begin{split}
       \frac{\partial b_-^i}{\partial t} =& \frac{1}{2}[(b_+^j\nabla_j - b_-^j\nabla_j)b_+^i- (b_+^j\nabla_j \\
       &+b_-^j\nabla_j) b_-^i] + \frac{\hbar}{2m}\Delta b_+^i -\frac{\nabla^i\phi}{m}.
    \end{split}
\end{equation}
In summary, (\ref{fwdPDE}) and (\ref{bwdPDE}), or equivalently, (\ref{fwddynamics}) and (\ref{bwddynamics}), recover Nelson's theory. From them, the Schr\"{o}dinger equation (\ref{SE}) can be derived through the similar set of mathematical transformations $\nabla^iR=\frac{m}{2\hbar}(b_+^i-b_-^i)$, $\nabla^i S=\frac{m}{2\hbar}(b_+^i+b_-^i)$, and $\psi=e^{R+iS}$.

The variational method presented here can be considered as an extension of Yasue's variational method~\cite{Yasue}. In fact, Yasue's variational method is a special case if we define total Lagrangian functional as $J_{ab} = (J_{ab}^++J_{ab}^-)/2 = \int E[L_Y]dt + \beta H_{rel}$. But the total relative entropy $H_{rel}$ gives no constraint as mentioned in Section \ref{subsec:relH}. This is why there is no constraint term in Yasue's variational method. 

It is worth to note in our derivation of Nelson's equations, and consequently, the Schr\"{o}dinger equation, we only rely on a limited number of definitions and assumptions. Specifically, there are three key definitions: 1.) The definitions of forward and backward derivatives (\ref{Derivative}), and forward and backward mean velocities; 2.) The identity (\ref{osmotic}), which in turn is derived from definitions of forward and backward mean velocities and Bayes' Theorem; 3.) Definition of Lagrangian $L_Y^{\pm}$ for both forward and backward path configurations. The important assumptions include: 1.) The system is constantly performing Markov diffusion; 2.) The variation principle of extremizing the action with constraints on relative entropy; 3.) The Lagrangian multiplier $\beta$ is set to be the Planck constant $\hbar$. The third assumption is non-trivial and is an important feature of stochastic mechanics. Its possible justification has been discussed in Ref.~\cite{Cetto, Pena}. The rest of the definitions and assumptions can be naturally understood from classical physics and probability theory. The fact that the Schr\"{o}dinger equation can be derived from this limited list of definitions and assumptions is striking. It manifests the original goal of stochastic mechanics to explain quantum mechanics from classical physics and principles as much as possible\footnote{It is likely that the list of definitions and assumptions needs to be expanded in order to recover other parts of quantum theory such as quantum measurement or quantum entanglement. These are future research topics.}

In addition, the variation method presented here gives richer physics since it derives PDEs for both forward and backward path configurations. The relative entropy constraint requires that the information encoded on the forward and backward path configuration is identical, indicating some kinds of information symmetry. The methodological implication of our approach will be discussed further in Section \ref{sec:conclusion}. Another interesting feature here is that there is no need to depend on the forward and backward Fokker-Planck equations, which are needed in Nelson's original derivation~\cite{Nelson}, and earlier variation approaches~\cite{Yasue, Guerra, Pavon, Caticha} in order to derive the Schr\"{o}dinger equation. To the contrary, in our variational approach, one can actually derive the Fokker-Planck equations from (\ref{CombinedPDE}) and (\ref{osmotic}), as shown in Appendix \ref{AppendixD}.

\subsection{Variations with Fisher Information}
\label{subsec:var2}
Now we turn to the question on why the Guerra version of Lagrangian (\ref{L_G}) can also lead to the same Nelson theory. We wish to derive the theory using the variation method similar to that in Section \ref{subsec:var1}, i.e., through a functional that combines both action and certain information measure of the diffusion process. Clearly such information measure cannot be the relative entropy, so our first step is to search what the suitable information measure might be. It turns out that the answer is related to Fisher information.

Traditionally, Fisher information is defined to measure the amount of information that an probability distribution of random variable $\mathbf{x}$ carries about an observable parameter $\theta$. Let $f(\mathbf{x}, \theta)$ the probability density function for $\mathbf{x}$ conditioned on $\theta$, the Fisher information is defined as~\cite{Frieden}
\begin{equation}
    \label{Fisher}
    I(\theta) = \int (\frac{\partial}{\partial\theta}f(\mathbf{x}, \theta))^2f(\mathbf{x}, \theta) d\mathbf{x}.
\end{equation}
If we are interested in the Fisher information about the observable of position for a probability density function, we can set the parameter $\theta$ as the position variable itself, i.e., let $\theta=x^i$ and $f(\mathbf{x}, \theta)=\rho(\mathbf{x})$, the Fisher information can be rewritten as~\cite{Reginatto, Parwani}
\begin{equation}
    \label{FI}
    I = \int \rho(\mathbf{x})(\nabla ln\rho(\mathbf{x}))^2 d\mathbf{x} = E[(\nabla ln\rho(\mathbf{\xi}))^2]
\end{equation}
Since $\nabla ln\rho = \mathbf{u}/\nu$ and $\nu=\hbar/2m$, $I$ can be rewritten as\footnote{It is worth to note that the Fisher information defined here is also related to the Bohm quantum potential. Let $Q=-\hbar^2\Delta^i\Delta_i\sqrt{\rho}/(2m\sqrt{\rho})$ be the Bohm quantum potential, it can be shown that~\cite{Carroll} $E[Q]=\hbar^2 I/(8m)$. \added{Bohm potential is considered to be nonlocal. The non-locality issue of stochastic mechanics is further discussed in Section IV.D.}}
\begin{equation}
    \label{FI2}
    I(\mathbb{\xi})= E[u^iu_i]/\nu^2.
\end{equation}
We further define the Fisher information production along the diffusion path $\mathbf{\xi}(t)$ from $t_a$ to $t_b$ as
\begin{equation}
    \label{FI2}
    \mathcal{I}_{ab} = \int_{t_a}^{t_b} \nu I(\mathbb{\xi}(t)) dt = \frac{1}{\nu}\int_{t_a}^{t_b}E[u^iu_i]dt.
\end{equation}
Constant $\nu$ is multiplied to $I$ so that $\mathcal{I}_{ab}$ is dimensionless.  With help of Theorem \ref{relativeH}, we have the following theorem.
\begin{theorem} 
\label{FisherTheorem}
For a Markov diffusion process $\mathbf{\xi}(t)$ from $t_a\to t_b$, the Fisher information production can be written as
\begin{equation}
\label{theo2}
\begin{split}
     \mathcal{I}_{ab}^+ &= H_b - H_a - \int_{t_a}^{t_b} E[\nabla_i b_+^i]dt, \\
     \mathcal{I}_{ab}^- &= H_a - H_b + \int_{t_a}^{t_b} E[\nabla_i b_-^i]dt,
\end{split}
\end{equation}  
\end{theorem}
Proof of Theorem 2 is given in Appendix \ref{AppendixE}. 

The Guerra and Morato version of Lagrangian are given as~\cite{Guerra}
\begin{equation}
    \label{L_G2}
    \begin{split}
    L_G^+(\mathbf{\xi, b_+}) &= \frac{1}{2}m(b_+)_ib_+^i + \frac{\hbar}{2}\nabla_ib_+^i - \phi(\xi) \\
    L_G^-(\mathbf{\xi, b_-}) &= \frac{1}{2}m(b_-)_ib_-^i - \frac{\hbar}{2}\nabla_ib_-^i - \phi(\xi).
    \end{split}
\end{equation}
The corresponding actions can be defined as
\begin{equation}
    \label{G_Action}
    \begin{split}
    \mathcal{A}_{ab}^+ &= \int_{t_a}^{t_b} E [L_G^+(\mathbf{\xi, b_+})]dt \\
    \mathcal{A}_{ab}^- &= \int_{t_a}^{t_b} E [L_G^-(\mathbf{\xi, b_-})]dt .
    \end{split}
\end{equation}
However, it is verified~\cite{Nelsonbook} that $E [L_G^+(\mathbf{\xi, b_+})]=E [L_G^-(\mathbf{\xi, b_-})] = E[\frac{1}{2}mv^iv_i-\frac{1}{2}m u^iu_i- \phi(\xi)]$. Thus the two expressions in (\ref{G_Action}) are equivalent. Furthermore, one can observe that the difference between $\mathcal{A}_{ab}^+$, defined in (\ref{G_Action}), and $A_{ab}^+$, defined in (\ref{YAction}), is a term related to $\int E[\nabla_i b_+^i]dt$. This indicates that the difference is related to the Fisher information production $\mathcal{I}_{ab}^+$. Thus, instead of minimizing the actions, we seek to minimize the actions and Fisher information production together, with the same relative entropy constraints. With this consideration, we define the Lagrangian functional as
\begin{equation}
    \label{LF2}
    \begin{split}
    \mathcal{J}_{ab}^+ &= (\mathcal{A}_{ab}^+ + \alpha \mathcal{I}_{ab}^+) - \beta H(\rho_+\|\rho_-) \\
    \mathcal{J}_{ab}^- &= (\mathcal{A}_{ab}^- + \alpha \mathcal{I}_{ab}^-) - \beta H(\rho_-\|\rho_+) 
    \end{split}
\end{equation}
where $\beta$ is the Lagrangian multiplier. Here, we seek to minimize the combination of actions and Fisher information production with the relative entropy constraints for both forward and backward paths, respectively. Substituting (\ref{theo1_2}), (\ref{theo2}), (\ref{G_Action}) into (\ref{LF2}), we have
\begin{equation}
    \begin{split}
    \mathcal{J}_{ab}^+ =& \int_{t_a}^{t_b} E[\frac{1}{2}m(b_+)_ib_+^i - \phi(\xi) +(\frac{\hbar}{2}-\alpha)\nabla_ib_+^i \\ &+\frac{\beta}{2}\nabla_i(b_+^i + b_-^i)]dt
     +(\alpha-\beta)(H_b - H_a).
    \end{split}
\end{equation}
The Lagrangian multiplier has been set as $\beta=\hbar$ in Section \ref{subsec:var1}. Let $\alpha=\hbar/2$, $\mathcal{J}_{ab}^+$ is simplified to\footnote{As the case of setting $\beta=\hbar$, the assumption of is $\alpha=\hbar/2$ also non-trivial. Its possible justification has been discussed in Ref. [29, 30].}
\begin{equation}
    \begin{split}
    \mathcal{J}_{ab}^+ =& \int_{t_a}^{t_b} E[\frac{1}{2}m(b_+)_ib_+^i - \phi(\xi)  \\ 
    & +\frac{\hbar}{2}\nabla_i(b_+^i + b_-^i)]dt
    -\frac{\hbar}{2}(H_b - H_a).
    \end{split}
\end{equation}
$\mathcal{J}_{ab}^+$ is different from $J_{ab}^+$ only by a constant term $\frac{\hbar}{2}(H_b - H_a)$. Applied the stochastic calculus described in Section \ref{calculus} by varying the diffusion process $\mathbf{\xi}(t)$ to be $\mathbf{\xi}(t)+\mathbf{z}(t)$ with $\mathbf{z}(t_a)=\mathbf{z}(t_b)=0$, the variation of $\mathcal{J}_{ab}^+$ can be calculated as
\begin{equation}
    \label{LF4}
    \delta\mathcal{J}_{ab}^+ = \int_{t_a}^{t_b} E[m(b_+)_i\delta b_+^i - \nabla_i\phi z^i + \frac{\hbar}{2}\nabla_i(\delta b_+^i + \delta b_-^i )]dt.
\end{equation}
Comparing (\ref{LF4}) to (\ref{J2}), we see $\delta\mathcal{J}_{ab}^+$ is the same as $\delta J_{ab}^+$. The subsequent calculations for (\ref{J4})-(\ref{J5}) are applicable here, and resulting the same PDE (\ref{fwdPDE}) for the dynamics of the forward path. Similarly, one can derive the expression for $\mathcal{J}_{ab}^-$
\begin{equation}
    \begin{split}
    \mathcal{J}_{ab}^- =& \int_{t_a}^{t_b} E[\frac{1}{2}m(b_-)_ib_-^i - \phi(\xi) \\ 
    & -\frac{\hbar}{2}\nabla_i(b_+^i + b_-^i)]dt
     +\frac{\hbar}{2}(H_b - H_a).
    \end{split}
\end{equation}
Again, $\mathcal{J}_{ab}^-$ is different from $J_{ab}^-$ only by a constant $\frac{\hbar}{2}(H_b - H_a)$. Variations on both functional give the same PDE (\ref{bwdPDE}) for the backward path. Thus, even though $\mathcal{J}_{ab}^{\pm}$ and $J_{ab}^{\pm}$ are defined quite differently, the variations over both Lagrangian functional converge to the same outcomes.

Essentially, compared to $L_Y^{\pm}$, the Guerra and Morato version of Lagrangian $L_G^{\pm}$ includes a term related to Fisher information production. We define the corresponding Lagrangian functional $\mathcal{J}_{ab}^{\pm}$ by reversing the effect of this term. Then, applying the same variation method with the relative entropy constraint, we can also recover Nelson's theory.

\begin{table*}[ht]
\caption{Choices of Lagrangian and Correspondent Variation Constraints}
\label{tab:1}  
\renewcommand*{\arraystretch}{1.4}
\begin{tabular}{|m{6cm}|m{5cm}|m{5cm}|}
\hline
 \textbf{Choice of Lagrangian} & \textbf{Variation Method} & \textbf{Augmented Action} \\
\hline
$L_Y^{\pm}=\frac{1}{2}m(b_{\pm})_ib_{\pm}^i - \phi$ & Add relative entropy constraint & $J_{ab}^{\pm} = A_{ab}^{\pm} - \beta H(\rho_{\pm}\|\rho_{\mp})$\\ 
\hline
$L_G^{\pm}=\frac{1}{2}m(b_{\pm})_ib_{\pm}^i \pm \frac{\hbar}{2}\nabla_ib_{\pm}^i - \phi$ & Add Fisher information production, and relative entropy constraint & $\mathcal{J}_{ab}^{\pm} = (\mathcal{A}_{ab}^{\pm} +  \mathcal{I}_{ab}^{\pm}) - \beta H(\rho_{\pm}\|\rho_{\mp}) $\\
\hline
$L_E^{\pm}=\frac{1}{2}m(b_{\pm})_ib_{\pm}^i \pm \frac{\hbar}{2}\nabla_i(b_+^i+b_-^i) - \phi$ & No constraint & $\mathbb{A}_{ab}^{\pm}$\\
\hline
\end{tabular}
\end{table*}
\subsection{Effective Lagrangian}
\label{subsec:Lagrangian}
If both Lagrangians $L_Y^{\pm}$, defined in (\ref{LYdir}), and $L_G^{\pm}$, defined in (\ref{L_G2}), can lead to the same PDEs for the forward and backward paths, one may ask why there can be multiple choices of Lagrangian for the same physical process. In classical mechanics, one typically writes down the kinetic energy $K$ and define Lagrangian as $L=K-\phi$ where $\phi$ is the potential energy. But in stochastic mechanics, there is no first principle that can guide the definition of Kinetic energy, since there are multiple variables related to velocity.

Further complication is that the Lagrangian functional in the variation method consists the action, which is a functional of the Lagrangian, and the relevant constraints from information measures. Technically once can choose an effective Lagrangian that factors in the constraints, as long as the variation on the Lagrangian functional without the constraint gives the desired dynamics PDEs. For instance, define
\begin{equation}
\label{Leff}
\begin{split}
    L_{E}^+ &= \frac{1}{2}m(b_+)_ib_+^i + \frac{\hbar}{2}\nabla_i(b_+^i+ b_-^i) - \phi \\
    L_{E}^- &= \frac{1}{2}m(b_-)_ib_-^i - \frac{\hbar}{2}\nabla_i(b_+^i+ b_-^i) - \phi.
    \end{split}
\end{equation}
The corresponding actions are 
\begin{equation}
    \label{E_Action}
    \begin{split}
    \mathbb{A}_{ab}^+ &= \int_{t_a}^{t_b} E [L_E^+]dt \\
    \mathbb{A}_{ab}^- &= \int_{t_a}^{t_b} E [L_E^-]dt .
    \end{split}
\end{equation}
Then, by applying the same variational method in previous section to minimize the actions $\mathbb{A}_{ab}^{\pm}$ without any other constraints, one can obtain the same PDEs as (\ref{fwdPDE}) and (\ref{bwdPDE}).

Table 1 summaries that for the three sets of Lagrangian, one can apply the variation calculus to derive the same Nelson's theory by combining with correct choices of Fisher information production and (or) relative entropy constraint.

By taking the absolute expectation of these three forms of Lagrangian, we may obtain additional insight. Recalled that $L_Y=(L_Y^++L_Y^-)/2$, one can verify that $L_Y=\frac{1}{2}mv^iv_i+\frac{1}{2}mu^iu_i-\phi$. Using identity (\ref{id}), we can express the expectations of the three forms of Lagrangian in terms of $L_Y$, $\mathbf{v}$ and $\mathbf{u}$,
\begin{equation}
    \label{LExp}
    \begin{split}
    E[L_Y^{\pm}]&=E[L_Y \pm mv^iu_i] \\
    E[L_G^{\pm}]&=E[L_Y-mu^iu_i] \\
    E[L_E^{\pm}]&=E[L_Y].
    \end{split}
\end{equation}
Yasue's initial variational approach using $L_Y]$ as the Lagrangian does not require constraint~\cite{Yasue}. (\ref{LExp}) shows that $E[L_E^{\pm}]=E[L_Y]$. Chosen this form of Lagrangian, the variational method does not require constraint to derive the Nelson theory, consistent with Yasue's approach. However, using $L_E^{\pm}$ as the Lagrangian gives the advantage of distinguishing the dynamics of forward and backward path configurations, which is missing in Yasue's approach. The difference between $E[L_Y^{\pm}]$ and $E[L_E^{\pm}]$ is the term $E[v^iu_i]$, which is related to the rate of entropy production. Thus, to use $L_Y^{\pm}$ as the Lagrangian, the relative entropy constraint is introduced in the variational method. On the other hand, the difference between $E[L_G^{\pm}]$ and $E[L_E^{\pm}]$ is the term $E[u^iu_i]$, which is related to Fisher information. Therefore, the Fisher information production is needed in the Lagrangian functional.

\section{Discussion and Conclusion}
\label{sec:conclusion}

\subsection{Degrees of Freedom}

\begin{figure*}
\begin{center}
\includegraphics[scale=3.9]{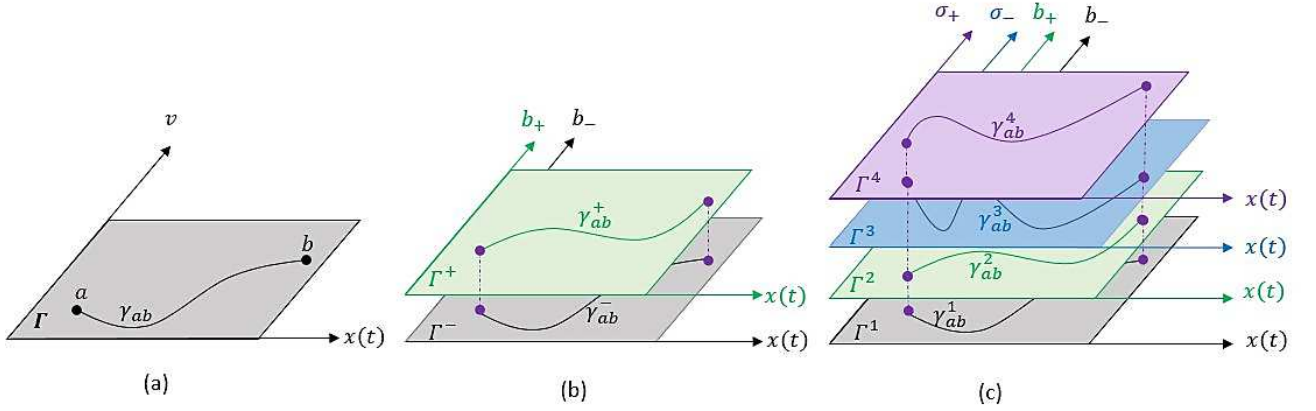}
\caption{(a) In classical mechanics, a point particle moves from point $a$ to point $b$ with a path configuration $\gamma_{ab}$ determined by least action principle. (b) In stochastic mechanics, diffusion of a point particle is described with forward and backward path configurations $\gamma^{\pm}_{ab}$. Each path configuration follows its own stochastic differential equation, but connected through the relative entropy constraint. Combing the two PDEs results in the Schr\"{o}dinger equation. (c) Conjecture: By introducing rotational degrees of freedom $\sigma^{\pm}$, there are four path configurations $\gamma^i$ in the phase spaces $\Gamma^i$ $(i=1,2,3,4)$. Can the variation approach developed here lead to the Dirac equation for spin once it is extended into the relativistic framework?}
\label{fig:2}       
\end{center}
\end{figure*}
In previous sections, we recast the theory of stochastic mechanics into two PDEs for forward and backward path configuration using the variation approach that combines the least action principle and information measure constraints. What are the physical implications of this derivation? 
The concept of forward and backward paths gives a physical picture when considering the dynamics of a diffusing particle in stochastic mechanics. Recalled that the term forward or backward does not refer to the direction of time. Instead, it just refers to the path configuration $\gamma_{ab}^{\pm}$ that are described by either the forward mean velocity and backward mean velocity, respectively. This implies that extra degrees of freedom are needed to completely describe the dynamics of the system. For the time being we assume the system under study is a point particle in a three-dimension space. As mentioned earlier, we can envision the phase space $\Gamma: \{(\mathbf{x}, \mathbf{b}_+, \mathbf{b}_-)\}$ comprises two classical phase subspaces $\Gamma^+: \{(\mathbf{x}, \mathbf{b}_+)\}$ and $\Gamma^-: \{(\mathbf{x}, \mathbf{b}_-)\}$, as shown in Fig.\ref{fig:2}b. Path configurations $\gamma_{ab}^{\pm}$ follow their own dynamics in $\Gamma^{\pm}$, respectively. But they are connected by the relative entropy constraint. The overall dynamics is described by the Lagrangian functional defined in (\ref{LF_Y}). Applying the variation method gives (\ref{fwdPDE}) for forward path configuration, and (\ref{bwdPDE}) for backward path configuration.  

The Schr\"{o}dinger equation essentially combines two non-linear PDEs into a linear PDE through mathematical transformation. In this perspective, the wave function in quantum mechanics is just a mathematical tool that makes the calculation much easier. But the underline physics is stochastic mechanics that essentially demands extra degree of freedom to completely describe a diffusing point particle.

By replacing the classical velocity with forward and backward mean velocity, one obtains two PDEs that leads to the Schr\"{o}dinger equation. We may naturally ask if this approach can be further extended. Although the assumption that the system is a point particle with both forward and backward mean velocity is a step forward compared to describing the system as a classical point particle with a single velocity, it is still an over simplification. Suppose we add two new degrees of freedom, $\sigma^+$ and $\sigma^-$, to describe the rotation of the system clockwise and counterclockwise, respectively. The total phase configuration is expanded to be $\Gamma: \{(\mathbf{x}, \mathbf{b}_+, \mathbf{b}_-, \sigma^+, \sigma^-)\}$. In a similar way, we envision the phase space $\Gamma$ comprises four classical phase subspaces $\Gamma^1: \{(\mathbf{x}, \mathbf{b}_+)\}$, $\Gamma^2: \{(\mathbf{x}, \mathbf{b}_-)\}$, $\Gamma^3: \{(\mathbf{x}, \sigma^+)\}$, and $\Gamma^4: \{(\mathbf{x}, \sigma^+)\}$, as shown in Fig.\ref{fig:2}c. There are four path configurations $\gamma^i (i=1,2,3,4)$ in these phase subspaces. We would expect that more constraints will be imposed, and the variational method can still be applied to give four PDEs, one for each path configuration. Will this lead to, or recover, the more advanced quantum theory, possibly the Dirac equation for spin? To this end, one will need to first construct a relativistic stochastic mechanics~\cite{Serva, Lindgren, Kuipers}, and also extend the variational approach in a relativistic setting. This is an interesting conjecture for future research.

Furthermore, other degrees of freedom can be added for a more sophisticated physical modeling of a quantum system.

\subsection{Probability for Indistinguishable Alternatives}
The constraint of zero relative entropy we impose in the variation principle essentially requires that the probabilities for both forward and backward paths are identical, i.e., $\rho_+=\rho_-$. By measuring the probability of a particle diffuses from point $a$ to point $b$, one cannot tell the underline paths are forward or backward paths. In other words, we can view the two path configurations as indistinguishable alternatives that are needed to complete the description of the diffusion dynamics. It is not clear if it is even possible to design an experiment to distinguish the two path configurations. But is clear that without a specific measurement, the forward and backward paths cannot be distinguishable. They are not exclusive alternatives.

Such similar idea was proposed in the Feynman path integral formulation of quantum mechanics~\cite{Feynman48}. In path integral, a particle can move from point $a$ to point $b$ through infinite number of alternative paths. These paths are indistinguishable but they should be all counted in order to have a complete description of the quantum behavior of the particle. The law for computing the probability of the particle moving from point $a$ to point $b$ is not just to simply add up the probability of each path. Instead, one adds up the probability amplitude for each path first, then takes the modulus square of the summation. According to Feynman~\cite{Feynman48, Feynman05}, the law of computing the probability is different from classical law because the alternatives are interfering alternatives, instead of exclusive alternative. The concept of interfering alternatives is unique to quantum mechanics.

In our formulation, there is no concept of probability amplitude. Thus, we cannot straightforwardly use the same law of calculating the probability as that in path integral. But there is still similarity in that the forward and backward paths are non-exclusive alternatives. The two paths interfere each other, as shown by the two PDEs (\ref{fwdPDE}) and (\ref{bwdPDE}). The probability of a particle diffuses from point $a$ to point $b$ is not simply the sum of the probabilities of each path. The characteristic of interfering alternative that is unique to quantum mechanics is manifested here, but in the context of stochastic process.

\subsection{Physical Variable versus Information Measure}
As pointed out in Section \ref{subsec:Lagrangian}, using the same stochastic calculus of variation, one can derive the Nelson theory, thus the Schr\"{o}dinger equation, from three different forms of Lagrangian. The action derived from each form of Lagrangian needs to be combined with appropriate constraints related to information measures in the variational method. Effectively, one can just construct a Lagrangian without being aware of the corresponding constraint, as long as the effective Lagrangian leads to the correct form of Euler-Lagrangian equation. This is possible because there is no clear principle to guide the construction the correct Lagrangian, and some of the constraints may not be obvious to recognize. This is the case for $L_E^{\pm}$. The terms in the Lagrangian come from proper choice of physical variables, such as kinetic energy and potential energy, or can come from information measure such as entropy production rate and Fisher information. In the case of stochastic mechanics, the Planck constant appears to be the Lagrangian multiplier that converts an information measure related term to act like physical variable term in the variation process. 

In both classical and quantum field theory (QFT), the first step of developing a field theory is to construct a proper Lagrangian density. This important first step is a creative one, allowing trial and error. As long as the Lagrangian density leads to the correct form of Euler-Lagrangian equation, it is accepted as an appropriate form. Many complicated Lagrangian density functions constructed in QFT comprise many terms, and some of the terms are introduced intuitively. If we extend the variation principle developed here for stochastic mechanics to field theory, we can reasonably ask the following question for future investigation: \textit{From stochastic mechanics perspective, are some of the terms in the Lagrangian density functions in classical or quantum field theory actually reflect certain information related constraints?}

At the philosophical level, the structure of Lagrangian developed here implies that a physical theory can embrace both ontological component and epistemic component. A physical theory essentially describes how well a physical phenomenon can be observed. Such observation echoes the ideas brought up by other authors previously~\cite{Wheeler, FriedenBook}.

\subsection{Comparison with the Earlier Variational Methods}
The variational principle presented here utilizes the stochastic calculus of variations from Yasue~\cite{Yasue} and significantly extends Yasue's results. In particular, we are able to derive the PDEs for both forward and backward paths. Yasue's derivation is just a special case in our formulation. Guerra and Morato~\cite{Guerra} proposed a different variation approach where the end point of the path varies. Again, they do not derive different PDEs for forward and backward path. Both Yasue and Guerra variation approaches do not involve information measures compared to our approach.

Ref.~\cite{Pavon} introduces a saddle-point entropy production principle, where one seeks to extremize the entropy production in the diffusion process with a constraint that essentially comes from the Fokker-Planck equation. In this works, we recognize the equation on the entropy production is actually a constraint on the relative entropy in the path space, and no need to depend on Fokker-Planck equation. Our approach is more intuitive and more importantly, we are able to derive PDEs for both forward and backward path, which is not shown in Ref.~\cite{Pavon}.

A more recent development of variation based entropic dynamics is described in Ref.~\cite{Caticha}, where the probability distribution is derived from a variation of relative entropy. Such probability distribution implies a trajectory of Brownian motion. Then, variation based on principle of conservation of energy, together with the Fokker-Planck equation, gives the Schr\"{o}dinger equation. This approach requires two variation processes, and does not give the dynamics of both forward and backward paths. Furthermore, the relative entropy defined in Ref.~\cite{Caticha} is very different from the relative entropy in the present works. \added{The method of variation with relative entropy was investigated in many other contexts. For instance, the original Schr\"{o}dinger Bridge Problem was later reformulated to be a problem of minimizing relative entropy~\cite{Follmer}, which leads to connection to the mass transportation theory~\cite{Leonard}. In such reformulation, the relative entropy is defined on the bridging path probability measure relative to a reference path probability measure. Variation over the dynamic probability measure to minimize the relative entropy gives to the solution of the Schr\"{o}dinger Bridge Problem~\cite{Leonard}. However, the way relative entropy is used in our formulation is very different. Here, the relative entropy is defined based on probability densities for both forward and backward path configurations, and used as a constraint in the least action principle.}

Variation on the combination of action and Fisher information production to derive the Schr\"{o}dinger equation was also studied in Refs.~\cite{Reginatto, Parwani}. However, the derivation was primarily mathematical rather than based on a physical dynamics model such as Brownian motion. Thus, it did not provide the dynamics of both forward and backward paths. However, the justifications in Ref.~\cite{Parwani} for choosing Fisher information in their variational method maybe well applicable in the stochastic variational approach here.

In summary, the novelty of the variational method proposed here comes from its capability of deriving PDEs from both forward and backward paths, and the fruitful interplay between actions and information measures in the Lagrangian functional. These results enhance our understanding of stochastic mechanics.

\subsection{Limitations}
The rigorousness of the derivation presented here depends on the stochastic calculus described by Yahsue~\cite{Yasue}. Thus, it is only as rigorous as the stochastic calculus can be. One limitation to point out is the calculation in (\ref{dJ1}). There the variation of $J_{ab}$ is calculated as $J_{ab}=E\int L(\mathbf{\xi}+\mathbf{z})dt - E\int L(\mathbf{\xi})dt = E\int (L(\mathbf{\xi}+\mathbf{z})-L(\mathbf{\xi}))dt $. Essentially it assumes that the variation due to the change of probability density along the path is negligible compared to the variation due to the change of integrand, so that the expectation operators are considered the same. A rigorous treatment to confirm this is desirable, even though this stochastic calculus has been successfully applied to derive Nelson's theory, Noether's theorem~\cite{Yasue}, the results in the present works, and in many other applications~\cite{Zambrini, Zambrini2, Cresson}. 

The origin of randomness of the diffusion process is not investigated here. One explanation is that the space metric tensor itself given stochastically with some appropriate distribution function~\cite{Kurihara}. A point particle in such stochastic metric space is described as Brownian motion. However, modeling a system as a point particle is an over simplification. As mentioned earlier, an intuitive extension is to introduce rotational degree of freedom into the variation, and investigate what new physical dynamics can be derived. \added{In addition, our construction here is limited to Markov diffusion process. The connection between quantum mechanics and a more generalized reciprocal stochastic process called Bernstein-Markov process was investigated in Ref.~\cite{Zambrini}, which confirms not only the connection between Bernstein process and Nelson's theory, but also the connection between Bernstein process and the imaginary time version of Schr\"{o}dinger equation. Whether the variational principle proposed in the present work can be applied to Bernstein reciprocal process is an interesting topic for further research.}

The stochastic mechanics has its own limitation to explain the locality problem. Nelson has described a scenario~\cite{Nelsonbook} that the behavior of a particle described by Markovian stochastic mechanics depends on the process on a second particle that is uncoupled and separated arbitrary away. Such scenario leads him to believe that stochastic mechanism is not a tenable physical theory to reflect reality~\cite{Nelsonbook}. \added{However, it is not clear whether stochastic mechanics should be regarded as a ``non-local” hidden variable theory that violates the Bell inequalities. Recent investigation shows that the derivation of Bell inequalities depends on three assumptions: outcome independence, statistical locality, and measurement independence~\cite{Hall}. Violation of Bell inequalities just means at least one of the three assumptions does not hold. The violation does not necessarily imply the theory is non-local. Instead it means the statistical correlation is non-separable~\cite{Hall2}. The non-locality issue that Nelson concerned on stochastic mechanics may well be due to the similar non-separability of statistical correlation that exhibits in the quantum correlation through entanglement.} 

There are other opened questions on how stochastic mechanics can explain quantum phenomena, readers are referred to the well known review papers~\cite{Goldstein, Nelson2} and a most recent recent paper~\cite{Kuipers}. One of the subtleties is that to what extend stochastic mechanics can rely on the classical probability theory to explain quantum mechanics specific phenomena. For instance, multi-time correlations cannot be straightforwardly calculated using classically probability theory since it predicts different results from quantum mechanics~\cite{Grabert}. Instead, to obtain the correct multi-time correlations, one should take account into consideration that after a measurement, the diffusion process is reset~\cite{Blanchard}. 

The challenges and limitations mentioned in this subsection will continue to inspire future researches to bring new physical insights into stochastic mechanics. The present works does not intend to address these opened issues in stochastic mechanics. But we believe the variational principle described here gives new insight that stochastic mechanics is coupled with information measure constraint \added{such as Fisher information that may give rise to the property of non-separability. Therefore,} it might explain the locality issue and entanglement phenomenon. This is currently under further investigation. 

\subsection{Conclusions}
A new variational principle is proposed here to derive the dynamics equations for the forward and backward paths, which when combined together, result in the non-relativistic Schr\"{o}dinger equation. According to this principle, appropriate Lagrangian must be chosen, together with constraints that are related to information measure such as relative entropy or Fisher information. The derivation method is based on the stochastic calculus. We show three different forms of Lagrangian can lead to the same Nelson theory. The advantages of this variational principle compared to others are clearly shown from its ability to resolve the issue of multiple Lagrangians, and the derivation of the Fokker-Planck equation as a side outcome instead of dependent on it when deriving the Sch\"{o}dinger equation.

The variational principle developed in this work not only mathematically recovers the Nelson stochastic mechanics and the non-relativistic Schr\"{o}dinger equation, but also brings new insights on the underlined physics. First,, the concept of forward and back paths provides a intuitive physical picture to consider the dynamics of a diffusing particle. It gives more insight on the difference between classical and quantum mechanics in terms of degree of freedom needed to describe a system completely. Methodologically, one can include new degrees of freedom to expand the theory to derive more advanced theory. It is natural to conjecture that this idea can be potentially generalized to derive the Dirac equation if rotational degrees of freedom are included in the stochastic variation. The constraint of zero relative entropy imposed to the least action principle shows that the forward and backward paths, although both are needed for a complete description of a diffusing particle, are indistinguishable through measurement. This echoes the idea of interfering alternative proposed in the path integral formulation of quantum mechanics. Finally, when constructing a Lagrangian, it is subtle to distinguish terms coming from physical variables such as kinetic energy and potential energy or from constraints on certain information measure such as relative entropy and Fisher information. It is intuitive to speculate that if the stochastic mechanics is extended to field theory, some of the Lagrangian terms in the field theory may turn out to actually reflect certain information constraints.

\begin{acknowledgements}
The author would like to thank the anonymous referees for their valuable comments, which help to clarify the physical implications of the present work, and the connection of the present work with the history of the stochastic mechanics.
\end{acknowledgements}

\section*{Data Availability Statement}
The data that support the findings of this study are available within the article.
 
\appendix

\section{Proof of Theorem \ref{relativeH}}
\label{AppendixA}
To prove Theorem \ref{relativeH}, one substitutes (\ref{fwdp}) and (\ref{bwdp}) into (\ref{fwdH}) and rearranges the terms inside the logarithm function,
\begin{equation}
    \label{PT1}
    \begin{split}
        &H(\rho_+\|\rho_-) \\
        &=\int \rho_+ln(\frac{\rho(\mathbf{x}_1)}{\rho(\mathbf{x}_n)}\times\frac{p(\mathbf{x}_2|\mathbf{x}_1)}{p(\mathbf{x}_1|\mathbf{x}_2)}\times...
        \times \frac{p(\mathbf{x}_n|\mathbf{x}_{n-1})}{p(\mathbf{x}_{n-1}|\mathbf{x}_n)}){\cal{D}}\mathbf{x} \\
        &=\int \rho_+ln(\frac{\rho(\mathbf{x}_1)}{\rho(\mathbf{x}_n)}){\cal{D}}\mathbf{x} + \sum_{i=1}^{n-1} \int \rho_+ ln(\frac{p(\mathbf{x}_{i+1}|\mathbf{x}_{i})}{p(\mathbf{x}_{i}|\mathbf{x}_{i+1})}){\cal{D}}\mathbf{x}
    \end{split}
\end{equation}
Expanding the first term, labeled as $T_1$, one gets
\begin{equation}
    \label{PT2}
    \begin{split}
    T_1 &=\int \rho_+ln\rho(\mathbf{x}_1){\cal{D}}\mathbf{x} - \int \rho_+ln\rho(\mathbf{x}_n){\cal{D}}\mathbf{x} \\
    &=\int \rho(\mathbf{x}_1)ln\rho(\mathbf{x}_1)d\mathbf{x}_1 \prod_{i=1}^n\int p(\mathbf{x}_{i+1}|\mathbf{x}_i)d\mathbf{x}_{i+1}\\
   & -\int ln\rho(\mathbf{x}_n)d\mathbf{x}_n\int\rho(\mathbf{x}_1) \prod_{i=1}^{n-1}p(\mathbf{x}_{i+1}|\mathbf{x}_i)d\mathbf{x}_{i}.
    \end{split}
\end{equation}
Recalled the following two identities,
\begin{equation}
    \label{probIDs}
    \begin{split}
        &\int p(\mathbf{x}_{i+1}|\mathbf{x}_i)d\mathbf{x}_{i+1}=1 \\
        &\int\rho(\mathbf{x}_i)p(\mathbf{x}_{i+1}|\mathbf{x}_i)d\mathbf{x}_{i}=\rho(\mathbf{x}_{i+1}),
    \end{split}
\end{equation}
$T_1$ is simplified as
\begin{equation}
    \label{PT3}
    \begin{split}
    T_1 &= \int \rho(\mathbf{x}_1)ln\rho(\mathbf{x}_1)d\mathbf{x}_1 - \int \rho(\mathbf{x}_n)ln\rho(\mathbf{x}_n)d\mathbf{x}_n \\
    &= H(\mathbf{x}_n) - H(\mathbf{x}_1).
    \end{split}
\end{equation}
By Bayes' Theorem (\ref{Bayes}), $p(\mathbf{x}_{i+1}|\mathbf{x}_{i})/p(\mathbf{x}_{i}|\mathbf{x}_{i+1}))=\rho(\mathbf{x}_{i+1})/\rho(\mathbf{x}_i)$. Substituting this into the second term in (\ref{PT1}), labeled as $T_2$, we have
\begin{equation}
    \label{PT4}
    \begin{split}
    T_2 =& \sum_{i=1}^{n-1}(\int \rho_+ln\rho(\mathbf{x}_{i+1}){\cal{D}}\mathbf{x} - \int \rho_+ln\rho(\mathbf{x}_i){\cal{D}}\mathbf{x} )\\
    =& \sum_{i=1}^{n-1}(\int \rho(\mathbf{x}_{i})p(\mathbf{x}_{i+1}|\mathbf{x}_{i})ln\rho(\mathbf{x}_{i+1})d\mathbf{x}_{i}d\mathbf{x}_{i+1}\\
    &-\int \rho(\mathbf{x}_{i})p(\mathbf{x}_{i+1}|\mathbf{x}_{i})ln\rho(\mathbf{x}_{i})d\mathbf{x}_{i}d\mathbf{x}_{i+1})\\
    =& \sum_{i=1}^{n-1}\int \rho(\mathbf{x}_{i})(\int p(\mathbf{x}_{i+1}|\mathbf{x}_{i})ln\rho(\mathbf{x}_{i+1})d\mathbf{x}_{i+1} \\
    & - ln\rho(\mathbf{x}_{i}))d\mathbf{x}_{i}\\
    =& \sum_{i=1}^{n-1} E[\int p(\mathbf{x}_{i+1}|\mathbf{x}_{i})ln\rho(\mathbf{x}_{i+1})d\mathbf{x}_{i+1} - ln\rho(\mathbf{x}_{i})]
    \end{split}
\end{equation}
$E[\cdot]$ is the absolute expectation. Recalled the definitions of forward derivative (\ref{Derivative}) and and the conditional expectation (\ref{Et}), denoted $\Delta t = t_{i+1} - t_i$,
\begin{equation}
    \label{PT5}
    \begin{split}
    T_2 =& \sum_{i=1}^{n-1} E[\frac{E_t[ln\rho(\mathbf{\xi}_{i+1})] - ln\rho(\mathbf{\xi}_{i}) }{\Delta t}]\Delta t \\
    = & \sum_{i=1}^{n-1} E[D_+ln\rho(\mathbf{\xi}_i)]\Delta t\\
    = & \int_{t_a}^{t_b} E[D_+ln\rho(\mathbf{\xi}(t))] dt.
    \end{split}
\end{equation}
In the last step, we take $\Delta t\to 0$. Substituting (\ref{PT3}) and (\ref{PT5}) back to (\ref{PT1}), we finally obtain
\begin{equation}
    H(\rho_+\|\rho_-) = H_b - H_a + \int_{t_a}^{t_b} E[D_+ln\rho(\mathbf{\xi}(t))] dt.
\end{equation}
Derivation of $H(\rho_-\|\rho_+)$ follows similar steps, except we need to use the definition of backward derivative in (\ref{Derivative}). Setting $\mathbf{x}_n=\mathbf{x}_b$, $\mathbf{x}_1=\mathbf{x}_a$ gives (\ref{theo1}) in Theorem \ref{relativeH}.

To prove (\ref{theo1_2}), one expands $D_+ln(\rho)$ in (\ref{PT5}) using (\ref{Df}),
\begin{equation}
    \label{PT6}
    T_2 = \int_{t_a}^{t_b} E[(\partial /\partial t + {b_+^i}\nabla_i +\nu\Delta)ln(\rho)] dt.
\end{equation}
The first term vanishes with the integration because
\begin{equation}
    \label{PT7}
    \begin{split}
    \int_{t_a}^{t_b} E[\frac{\partial}{\partial t}ln(\rho)]dt &= \int_{t_a}^{t_b}\int \frac{\partial\rho}{\partial t}d\mathbf{x}dt\\
    &=\int \{\rho(\mathbf{x}(t_b)) - \rho(\mathbf{x}(t_a))\}d\mathbf{x} \\
    & = 0.
    \end{split}
\end{equation}
Rewriting the second in (\ref{PT6}) using (\ref{osmotic}), i.e., $u_i=\nu\nabla_iln\rho$, one gets $E[b_+^i\nabla_iln\rho] = \frac{1}{\nu}E[b_+^iu_i]$. Let $f^i=b_+^i$ in (\ref{id}), one obtains $\frac{1}{\nu}E[b_+^iu_i]= - E[\nabla_ib_+^i]$. Finally, for the third term in (\ref{PT6}), $E[\nu\Delta ln(\rho)]=E[\nabla_iu^i]$. Substitute these identities into (\ref{PT6}),
\begin{equation}
    \label{PT8}
    \begin{split}
    \int_{t_a}^{t_b} E[D_+ln(\rho)] dt &= \int_{t_a}^{t_b} E[\nabla_i(u^i - b_+^i)]dt \\
    &= - \int_{t_a}^{t_b} E[\nabla_iv^i]dt\\
    &= - \frac{1}{2}\int_{t_a}^{t_b} E[\nabla_i(b_+^i + b_-^i)]dt.
    \end{split}
\end{equation}
Similarly, one can also verify 
\begin{equation}
    \label{PT9}
    \int_{t_a}^{t_b} E[D_-ln(\rho)] dt = - \frac{1}{2}\int_{t_a}^{t_b} E[\nabla_i(b_+^i + b_-^i)]dt.
\end{equation}
Substituting (\ref{PT8}) and (\ref{PT9}) into (\ref{theo1}), we have (\ref{theo1_2}). 

\section{Swapping order of taking expectation and integration}
\label{AppendixB}
In the definition of action in (\ref{YAction}), $E[\cdot]$ is understood as the absolute expectation along the path $\mathbf{x}_a \to \mathbf{x}_b$. Given the property of Markov process, the probability density is defined in (\ref{fwdp}). Thus,
\begin{equation}
    \label{B1}
    \begin{split}
        A^+_{ab} =& \int \rho_+ \int_{t_a}^{t_b} L_Y^+dt \cal{D}\mathbf{x} \\
        =& \int \rho(\mathbf{x}_1)p(\mathbf{x}_2|\mathbf{x}_1)p(\mathbf{x}_3|\mathbf{x}_2)...p(\mathbf{x}_n|\mathbf{x}_{n-1})\\
        &\times \sum^n_{i=1}L_Y^+(\mathbf{x}_i, t_i)\Delta t d\mathbf{x}_1d\mathbf{x}_2...d\mathbf{x}_n \\
        =&\sum^n_{i=1}\Delta t \int \rho(\mathbf{x}_1)p(\mathbf{x}_2|\mathbf{x}_1)p(\mathbf{x}_3|\mathbf{x}_2)...p(\mathbf{x}_n|\mathbf{x}_{n-1}) \\
        &\times L_Y^+(\mathbf{x}_i, t_i) d\mathbf{x}_1d\mathbf{x}_2...d\mathbf{x}_n.
    \end{split}
\end{equation}
For each of the integration term in the summation over $i$, we apply the identities (\ref{probIDs}) to the integration over $d\mathbf{x}_1d\mathbf{x}_2...d\mathbf{x}_n$ except $d\mathbf{x}_i$, it ends up with
\begin{equation}
    \label{B2}
    \begin{split}
    A^+_{ab} =& \sum^n_{i=1}\Delta t \int \rho(\mathbf{x}_i) L_Y^+(\mathbf{x}_i, t_i)d\mathbf{x}_i \\
    =&\sum^n_{i=1}\Delta t E[L^+_Y (t_i)]\\
    =&\int_{t_a}^{t_b} E[L^+_Y(t)] dt.
    \end{split}
\end{equation}
Here $E[\cdot]$ denotes the absolute expectation over the configuration at time $t$. Thus, the order of integration and expectation can be swapped, as long as the meaning of expectation is understood correctly. The same argument goes for $A_{ab}^-$.

\section{Proof of $\delta b^i_+=D_+z^i$}
\label{AppendixC}
First note that $z^i(t) = \xi'^i(t) - \xi^i(t)$. By the definition of derivative operator $D_+$ in (\ref{Derivative}),
\begin{equation}
    \begin{split}
        D_+z^i &= \lim_{dt\to 0^+} E_t [\frac{z^i(t+dt)-z^i(t)}{dt}] \\
        &=\lim_{dt\to 0^+} E_t [\frac{\xi'^{i}(t+dt)-\xi^{i}(t+dt)-\xi'^{i}(t)+\xi^{i}(t)}{dt}] \\
        &=\lim_{dt\to 0^+} E_t [\frac{\xi'^{i}(t+dt)-\xi'^{i}(t)}{dt} - \frac{\xi^{i}(t+dt)-\xi^{i}(t)}{dt}] \\
        &=D_+\xi'^i(t) - D_+\xi^i(t). 
    \end{split}
\end{equation}
But $D_+\xi'^i(t)=b'^i_+(\xi'(t))$ and $D_+\xi^i(t)=b^i_+(\xi'(t))$. Therefore, $D_+z^i = b'^i_+(\xi'(t)) - b^i_+(\xi(t)) = \delta b^i_+$. Similarly, one can verify that $D_-z^i=\delta b^i_-$.

\section{Derivation of Nelson's First Equation (\ref{Nelson1}) and Fokker-Planck Equations from (\ref{CombinedPDE})}
\label{AppendixD}
Given $\beta=\hbar$ and (\ref{osmotic}), $\beta\nabla^i ln(\rho)=2mu^i$. Applied (\ref{Df}) to $u^i$, the R.H.S of (\ref{CombinedPDE}) becomes
\begin{equation}
    \label{PT31}
    \begin{split}
    R.H.S &= 2m(2\frac{\partial}{\partial t} + (b_+^j+b_-^j)\nabla_j)u^i \\
    &= 4m(\frac{\partial u^i}{\partial t} + v^j\nabla_ju^i).
    \end{split}
\end{equation}
Applied (\ref{Df}) to $b^i_{\pm}$, the L.H.S of (\ref{CombinedPDE}) becomes
\begin{equation}
    \label{PT32}
    \begin{split}
    L.H.S &= m(\frac{\partial}{\partial t} + b_-^j\nabla_j - \nu\Delta)b_+^i -m(\frac{\partial}{\partial t} + b_+^j\nabla_j + \nu\Delta)b_-^i \\
    &= m(2\frac{\partial u^i}{\partial t} + b_-^j\nabla_jb_+^i-b_+^j\nabla_jb_-^i-2\nu\Delta v^i).
    \end{split}
\end{equation}
Equating (\ref{PT31}) and (\ref{PT32}), after simple algebra, one gets
\begin{equation}
    \label{PT33}
    \begin{split}
    \frac{\partial u^i}{\partial t} &= -2v^j\nabla_ju^i + \frac{1}{2}(b_-^j\nabla_jb_+^i-b_+^j\nabla_jb_-^i)-\nu\Delta v^i \\
    & = -2v^j\nabla_ju^i + v^j\nabla_ju^i- u^j\nabla_jv^i -\nu\Delta v^i \\
    & = - v^j\nabla_ju^i - u^j\nabla_jv^i - \nu\Delta v^i.
 \end{split}
\end{equation}
Since both $u^i$ and $v^i$ are gradients and the curl of a gradient is zero, we have $\nabla_ju^i = \nabla^i u_j$, $\nabla_jv^i = \nabla^i v_j$, so that
\begin{equation}
    \label{PT33}
    \begin{split}
    \frac{\partial u^i}{\partial t} &= - v^j\nabla^iu_j - u^j\nabla^i v^j - \nu\Delta v^i \\
    &= - \nabla^i v^ju_j - \nu\Delta v^i.
    \end{split}
\end{equation}
This is the same equation as (\ref{Nelson1}). 

We further derive the Fokker-Planck equations from (\ref{PT33}). To do this, note that $\Delta v^i = \nabla^j\nabla_j v^i = \nabla^i\nabla_j v^j$, we rewrite the R.H.S of (\ref{PT33}) as $-\nabla^i(v^ju_j + \nu\nabla_jv^j)$, and since $u^i=\nu\nabla^i ln\rho$, the L.H.S. of (\ref{PT33}) is $\partial u^i/\partial t = \nu\nabla^i (\partial ln\rho/\partial t)$. Thus,
\begin{equation}
    \label{PT34}
    \nabla^i(\nu\frac{\partial ln\rho}{\partial t})= -\nabla^i(v^ju_j + \nu\nabla_jv^j) 
\end{equation}
This gives
\begin{equation}
    \label{PT35}
    \begin{split}
    \nu\frac{\partial ln\rho}{\partial t}&= -v^iu_i - \nu\nabla_iv^i \\
    &= -\nu(\nabla_iln\rho)v^i - \nu\nabla_iv^i\\
    &= -\nu(\frac{\nabla_i\rho}{\rho}v^i + \nabla_iv^i)
    \end{split}
\end{equation}
Multiplying $\rho/\nu$ to both sides gives the continuity equation of probability density
\begin{equation}
    \label{PT36}
    \frac{\partial \rho}{\partial t} = -\nabla_i(v^i\rho).
\end{equation}
Substituting $v^i = b_+^i - u^i = b_+^i - \nu\nabla^iln\rho$ one gets the forward Fokker-Planck equation
\begin{equation}
    \label{PT37}
    \frac{\partial \rho}{\partial t} = -\nabla_i(b_+^i\rho) + \nu\Delta\rho.
\end{equation}
Substituting $v^i = b_-^i + u^i = b_-^i + \nu\nabla^iln\rho$ one gets the backward Fokker-Planck equation
\begin{equation}
    \label{PT38}
    \frac{\partial \rho}{\partial t} = -\nabla_i(b_-^i\rho) - \nu\Delta\rho.
\end{equation}

\section{Proof of Theorem \ref{FisherTheorem}}
\label{AppendixE}
Let $f^i=u^i$ in (\ref{id}), we get $\frac{1}{\nu}E[u_iu^i] = -E[\nabla_iu^i]$. Substitute this into (\ref{FI2}),
\begin{equation}
    \label{PT21}
    \mathcal{I}_{ab} = - \int_{t_a}^{t_b} E[\nabla_iu^i] dt.
\end{equation}
Since $u^i = b_+^i - v^i$,
\begin{equation}
    \label{PT22}
    \begin{split}
    \mathcal{I}_{ab} &= - \int_{t_a}^{t_b} E[\nabla_ib_+^i] dt + \int_{t_a}^{t_b} E[\nabla_iv^i]dt\\
    &=- \int_{t_a}^{t_b} E[\nabla_ib_+^i] dt + \frac{1}{2}\int_{t_a}^{t_b} E[\nabla_i(b_+^i+b_-^i)]dt
    \end{split}
\end{equation}
From (\ref{theo1_2}) in Theorem \ref{relativeH} and Corollary \ref{cor1}, the second term equals $H(\mathbf{x}_b) - H(\mathbf{x}_a)$. Therefore,
\begin{equation}
    \label{PT23}
    \mathcal{I}_{ab} = H(\mathbf{x}_b) - H(\mathbf{x}_a) - \int_{t_a}^{t_b} E[\nabla_ib_+^i] dt.
\end{equation}
One can also substitute $u^i = v^i-b_-^i$ into (\ref{PT21}) and get
\begin{equation}
    \label{PT24}
    \begin{split}
    \mathcal{I}_{ab} &= \int_{t_a}^{t_b} E[\nabla_ib_-^i] dt - \int_{t_a}^{t_b} E[\nabla_iv^i]dt\\
    &=\int_{t_a}^{t_b} E[\nabla_ib_-^i] dt - \frac{1}{2}\int_{t_a}^{t_b} E[\nabla_i(b_+^i+b_-^i)]dt \\
    &=H(\mathbf{x}_a) - H(\mathbf{x}_b) + \int_{t_a}^{t_b} E[\nabla_ib_-^i] dt.
    \end{split}
\end{equation}
Both (\ref{PT23}) and (\ref{PT24}) are equivalent expression for the Fisher information production, but we label $\mathcal{I}_{ab}$ as $\mathcal{I}_{ab}^{\pm}$ in Theorem \ref{FisherTheorem} to reflect the dependency on the forward and backward mean velocity, respectively.


\begin{thebibliography}{}
%
%
\bibitem{Stanford} See, for example, the collection of quantum mechanics interpretations at Stanford University, https://plato.stanford.edu
\bibitem{Jammer74}M. Jammer, The Philosophy of Quantum Mechanics: The Interpretations of Quantum Mechanics in Historical Perspective. New York: Wiley-Interscience, (1974)
\bibitem{Goldstein}S. Goldstein, Stochastic Mechanics and Quantum Theory, J. Stat. Phys. 47, 645–667(1987)
\bibitem{Nelson2} E. Nelson, Review of Stochastic Mechanics, J. Phys. Conf. Ser. 361, 021011 (2012)
\bibitem{SBP}E. Schr\"{o}dinger. Sur la th\'{e}orie relativiste de l’\'{e}lectron et l’interpr\'{e}tation de la m\'{e}canique quantique. Ann. Inst. H. Poincar\'{e}, 2:269–310, (1932)
\bibitem{Follmer}H. F\"{o}llmer. Random fields and diffusion processes, in Ecole d’\'{e}t\'{e} de Probabilit\'{e}s de Saint-Flour XV-XVII-1985-87, volume 1362 of Lecture Notes in Mathematics. Springer, Berlin, (1988)
\bibitem{Cruz}A.B. Cruzeiro, L. Wu, J.C. Zambrini, Bernstein processes associated with a Markov process. In: Rebolledo R. (eds) Stochastic Analysis and Mathematical Physics. Trends in Mathematics, p41-71 Birkh\"{a}user, (2000)
\bibitem{Leonard}C. L\'{e}onard, A survey of Schr\"{o}dinger's problem and some of its connections with Optimal Transport. Discrete and Continuous Dynamical Systems, 34 (4), 1533-1574 (2014)
\bibitem{Fenyes} I. F\'{e}nyes, A Deduction of Schrödinger Equation, Acta Bolyaiana. 1 (5): ch. 2. (1946)
\bibitem{Nelson} E. Nelson, Derivation of the Schr\"{o}dinger Equation from Newtonian Mechanics, Phy. Rev. 150, 1079 (1966)
\bibitem{Nelsonbook}E. Nelson, Quantum Fluctuations, Princeton University Press (1985)
\bibitem{Yasue}K. Yasue, Stochastic Calculus of Variations, J. of Functional Analysis 41, 327-340 (1981)
\bibitem{Yasue2}K. Yasue, Quantum Mechanics and Stochastic Control Theory, J. Math. Phys. 22, 1010 (1981)
\bibitem{Zambrini}J. C. Zambrini, Stochastic dynamics: A Review of Stochastic Calculus of Variations, Int. J.  Theor. Phys. 24, 277–327 (1985)
\bibitem{Zambrini2}J. C. Zambrini, Variational Processes and Stochastic Versions of Mechanics, J. Math. Phys. 27, 2307 (1986)
\bibitem{Cresson}J. Cresson and S. Darses, Stochastic Embedding of Dynamical Systems, J. Math. Phys. 48, 072703 (2007)
\bibitem{Guerra}F. Guerra and L. I. Morato, Quantization of Dynamical Systems and Stochastic Control Theory, Phys. Rev. D, 1774-1786 (1983)
\bibitem{Pavon}M. Pavon, Hamilton's Principle in Stochastic Mechanics, J. of Math. Phys. 36 (12), 1995
\bibitem{Caticha}A. Caticha, Entropy Dynamics, Time, and Quantum Theory, J. Phys. A: Math. Theor. 44, 225303 (2011)
\bibitem{Santos}E. Santos, On a heuristic point of view concerning the motion of matter: From random metric to Schr\"{o}dinger equation, Phys. Lett. A 352: 49-54 (2006)
\bibitem{Kurihara}Y. Kurihara, Stochastic Metric Space and Quantum Mechanics, J. Phys. Comm. 2 (2018)
\bibitem{Pena2}L. de la P\~{e}na and A. M. Cetto, The quantum dice. An introduction to stochastic electrodynamics. Kluwer Academic Publishers, Dordrecht, (1996)
\bibitem{Santos2}E. Santos, Towards a stochastic interpretation of quantum physics, arXiv:1205.0916 (2012)
\bibitem{Serva}M. Serva, Relativistic Stochastic Processes Associated to Klein-Gordon Equation, Annales de l'I.H.P. Physique Th\'eorique 49, 415-432 (1988)
\bibitem{Lindgren}J. Lindgren, and J. Liukkonen, Quantum Mechanics Can Be Understood Through Stochastic Optimization on Spacetimes, Scr. Rep. 9:19984 (2019)
\bibitem{Kuipers} F. Kuipers, Stochastic Quantization on Lorentzian Manifolds, arXiv:2101.12552 (2021)
\bibitem{Wheeler} J. A. Wheeler, Time Today. In: Physical Origins of Time Asymmetry, Cambridge University Press, Cambridge UK (1994)
\bibitem{FriedenBook}B. R. Frieden, Science From Fisher Information, Cambridge University Press, Cambridge UK (2004)
\bibitem{Frieden}B. R. Frieden, Fisher Information as the Basis for the Schr\"{o}dinger Wave Equation, American J. Phys. 57, 1004 (1989)
\bibitem{Reginatto}M. Reginatto, Derivation of the equations of nonrelativistic quantum mechanics using the principle of minimum Fisher information, Phys. Rev. A 58, 1775 (1998)
\bibitem{Parwani}R. Parwani, Information Measures for Inferring Quantum Mechanics, J. Phys. A 38, 6231 (2005)
\bibitem{Carroll}R. Carroll, On the Emergence Theme of Physics, World Scientific, 2010, ISBN 981-4291-79-X, Chapter 1.
\bibitem{Cetto}A. M. Cetto, L. de la Pe\~{n}a, and A. Vald\'{e}s-Hern\'{a}ndez, Specificity of the Schr\"{o}dinger equation, Quantum Stud.: Math. Found. 2:275–287 (2005)
\bibitem{Pena}L. de la Pe\~{n}a, A. M. Cetto, and A. Vald\'{e}s-Hern\'{a}ndez, Connecting Two Stochastic Theories That Lead to Quantum Mechanics, Front. Phys. 8:162 (2020)
\bibitem{Fuchs} C. M. Caves, C. A. Fuchs, R. Schack, Quantum probabilities as Bayesian probabilities, Phys. Rev. A 65 22305 (2002). arXiv:quant-ph/0106133
\bibitem{Fuchs02}C. A. Fuchs, Quantum Mechanics as Quantum Information (and only a little more). arXiv:quant-ph/0205039, (2002)
\bibitem{Hoehn2014} P. A. H\"{o}hn, Toolbox for reconstructing quantum theory from rules on information acquisition, Quantum 1, 38 (2017)
\bibitem{Hoehn2015} P. A. H\"{o}hn, Quantum theory from questions, Phys. Rev. A 95, 012102 (2017),  arXiv:1511.01130v7
\bibitem{Yang2017}J. M. Yang, \href{https://doi.org/10.1038/s41598-018-31481-8}{A Relational Formulation of Quantum Mechanics, Sci. Rep. 8:13305 (2018),} arXiv:1706.01317
\bibitem{Feynman48}R. P. Feynman, Space-Time Approach to Non-Relativistic Quantum Mechanics, Rev. Mod. Phys. 20, 367, (1948)
\bibitem{Feynman05}R. Feynman and A. Hibbs, Quantum Mechanics and Path Integral. Emended by Styer, F., Dover Publications, New York (2005)
\bibitem{Grabert} H. Grabert, P. H\"{a}nggi, and P. Talkner, Is quantum mechanics equivalent to a classical stochastic process? Phys. Rev. A 19, 2440 (1979)
\bibitem{Blanchard} P. Blanchard, S. Golin, and M. Serva, Repeated Measurements in Stochastic Mechanics, Phys. Rev. D 34, 3732 (1986)
\bibitem{Nelson00} M. A. Nielsen and I. L. Chuang, Quantum computation and quantum information. Cambridge University Press, Cambridge UK (2000)
\bibitem{Hall}M. Hall, Relaxed Bell inequality and Kochen-Specker theorems. Phys. Rev. A 84, 022102 (2011)
\bibitem{Hall2}M. Hall, The significance of measurement independence for Bell inequalities and locality, In: At the Frontier of Spacetime, ed. T. Asselmeyer-Maluga, Springer, Switzerland, 189-204 (2016)
\end{thebibliography}


\end{document}